\documentclass[12pt]{article}
\usepackage[dvips]{graphicx}
\usepackage[tight]{subfigure}

\def\be{\begin{equation}}
\def\ee{\end{equation}}
\def\bea{\begin{eqnarray}}
\def\eea{\end{eqnarray}}
\def\beaN{\begin{eqnarray*}}
\def\eeaN{\end{eqnarray*}}
\def\ed{\end{document}}
\def\bit{\begin{itemize}}
\def\eit{\end{itemize}}

\def\sig{\sigma}

\def\lam{\lambda}

\def\Del{\Delta}

\def\k{\kappa}
\def\alf{\alpha}

\def\di{\partial}

\def\Lix{\pounds_\xi}

\def\half{{\textstyle{1 \over 2}}}
\def\~{\tilde}
\def\lag{{{\cal L}}}
\def\m{\label}
\def\l{\left}
\def\r{\right}

\def\goto{\rightarrow}

\def\const{\rm const}
\def\diag{\rm diag}

\def\sA{\stackrel{\bullet}{A}}
\def\sR{\stackrel{\bullet}{R}}
\def\sG{\stackrel{\bullet}{\Gamma}}
\def\cG{\stackrel{\circ}{\Gamma}}
\def\sS{\stackrel{\bullet}{S}}
\def\ssJ{\stackrel{\star}{J}}
\def\sD{\stackrel{\bullet}{\cal D}}
\def\sJ{\stackrel{\bullet}{J}}

\def\sK{\stackrel{\bullet}{K}}
\def\sT{\stackrel{\bullet}{T}}
\def\sL{\stackrel{\bullet}{\lag}}
\def\cN{\stackrel{\circ}{\nabla}}
\def\cA{\stackrel{\circ}{A}}
\def\cR{\stackrel{\circ}{R}}
\def\sU{\stackrel{\bullet}{\cal U}}
\def\MU{\stackrel{\scriptscriptstyle M}{\cal U}}
\def\MM{\stackrel{\scriptscriptstyle M}{\cal M}}
\def\MJ{\stackrel{\scriptscriptstyle M}{\cal J}}
\def\Mt{\stackrel{\scriptscriptstyle M}{\theta}}

\def\sM{\stackrel{\bullet}{\cal M}}
\def\scJ{\stackrel{\bullet}{\cal J}}
\def\sK{\stackrel{\bullet}{K}}
\def\stheta{\stackrel{\bullet}{\theta}}

\begin{document}

\centerline{\bf CONSERVED CURRENTS AND SUPERPOTENTIALS}
\centerline{\bf IN TELEPARALLEL EQUIVALENT OF GR}
\smallskip
\smallskip

\centerline{\it  E.D.Emtsova$^{1,2}$, A.N.Petrov$^{1}$, A.V.Toporensky$^{1,3}$}
\centerline{${^1}${\it Sternberg
Astronomical institute MV Lomonosov State university}}
\centerline {\it
 Universitetskii pr., 13, Moscow, 119992,
RUSSIA}

\centerline{${^2}$ \it Physical Faculty, Lomonosov Moscow State University, Moscow 119991, Russia}

\centerline{${^3}$ \it Kazan Federal University, Kremlevskaya 18, Kazan, 420008, Russia}

\centerline{e-mail: ed.emcova@physics.msu.ru}
\centerline{e-mail: alex.petrov55@gmail.com}
\centerline{e-mail: atopor@rambler.ru}
\smallskip
fax: +7(495)9328841

\smallskip

\begin{abstract}
 We study the Teleparallel Equivalent of General Relativity (TEGR) with Lagrangian that includes the flat (inertial) spin connection and that is evidently invariant with respect to local Lorentz rotations. Applying directly the Noether theorem, we construct new expressions for conserved currents and related superpotentials. They are covariant both under coordinate transformations and local Lorentz rotations, and allow us to construct well defined conserved charges, unlike earlier approaches. The advantage is achieved by an explicit  presence of a displacement vector in the new expressions that can be interpreted as a Killing vector, as a proper vector of an observer, etc. The new expressions are used to introduce a  principle for definition of an inertial spin connection that is undetermined one in the TEGR from the start. Theoretical results are applied to calculate mass for the Schwarzschild black hole and densities of conserved quantities for freely falling observers both in Friedmann-Lema\^itre-Robertson-Walker world of all the three signs of curvature and in (anti-)de Sitter space.
\end{abstract}

\section{Introduction}
\setcounter{equation}{0}

Last years the interest to Teleparallel Equivalent of General Relativity (TEGR) and its modifications, like $f(T)$-, $f(T_{ax}, T_{ten},T_{vec})$-theories and others, arises significantly;
see, for example, book \cite{Aldrovandi_Pereira_2013}, reviews \cite{Maluf,REV_2018} and numerous references therein. The advantage of such theories is that their equations are of the second order only, unlike other numerous modifications of GR in 4 dimensions. Among teleparallel theories the most developed one is TEGR, see \cite{Aldrovandi_Pereira_2013}. In spite of evident successes a construction of conservation laws and conserved quantities in TEGR feels big difficulties. Particularly these problems are discussed in the book \cite{Aldrovandi_Pereira_2013}, in more detail they are presented in presentation \cite{Krssak} by Martin Kr\v s\v s\'ak. Summating results in previous studies of many authors, he considers conserved energy-momentum complexes and related superpotentials.

The Kr\v s\v s\'ak requirements for constructing energy-momentum in TEGR are as follows. It has to 1) be of the first derivatives only, 2) be covariant with respect to both coordinate transformations and local Lorentz rotations, 3) permit to construct global (integral) conserved quantities, or conserved charges, 4) be symmetric and 5) be trace-free. We do not consider the two last requirements as so important. First, it is well known that canonical energy-momentum in a classical field theory is not symmetrical in general, however, it is not a problem for constructing all necessary conserved quantities, see Chapter 1 in the book \cite{Petrov_KLT_2017}. Second, the trace-free condition is rather in analogy with the massless electrodynamics. However, we would not provide this analogy as physically so important. Indeed, the electrodynamic field is considered as propagated on fixed or dynamic background spacetime, whereas the gravitational field presents  spacetime itself. One could consider perturbations of gravitational field in a given spacetime, however it is not the case that is considered by Kr\v s\v s\'ak \cite{Krssak}, and it is not the case that we consider in this paper.

The requirement 1) we consider as quite important one, and all variants of energy-momentum in TEGR satisfy it. The more interesting and important requirements by Martin Kr\v s\v s\'ak are the requirements 2) and 3). In \cite{Krssak}, it is remarked the problem that the known approaches give, on the one hand, well defined conserved charges expressed through well defined surface integrals, but one has only Lorentz non-covariant conserved energy-momentum. On the other hand, one can construct Lorentz covariant conserved energy-momentum, but then it  leads to undefined related conserved charges. In the paper, we resolve this problem, constructing conserved quantities in TEGR with making the use of the Noether theorem directly. As a result, the requirements 2) and 3) are satisfied simultaneously.

The paper is organized as follows. The next section \ref{Preliminaries} has a sense of preliminaries were we give all necessary information and outline of the outstanding problems of tetrad variants of general relativity. We present necessary definitions and related notations. We derive Moller's tetrad Lagrangian \cite{Moller_1961} that is not invariant with respect to local Lorentz rotations in whole. We derive also the Lagrangian given in the book \cite{Aldrovandi_Pereira_2013} that is Lorentz invariant because it
(unlike the Moller Lagrangian) contains very important structure - inertial spin connection.
Both of the Lagrangians lead to the same field equations. Both of them are classified as the TEGR Lagrangians.
However, for the sake of a definiteness, because we will compare consequences from both the Lagrangians, we will
call the presentation related to \cite{Moller_1961} as the tetrad Moller's one, whereas the presentation
given in \cite{Aldrovandi_Pereira_2013} as the covariant TEGR one. Following \cite{Aldrovandi_Pereira_2013}, in the framework of the covariant TEGR, we present conservation laws, energy-momentum complexes and superpotentials \cite{Aldrovandi_Pereira_2013}, and describe problems of these conserved quantities.

In section \ref{Noether}, in the framework of an {\em arbitrary} field theory, applying directly the Noether theorem with making
 the use of the diffeomorphism invariance, we derive conserved currents and related superpotentials of the most general form.

In section \ref{CL}, applying the results of section \ref{Noether}, we construct the conserved quantities both for the Moller \cite{Moller_1961} and for the covariant TEGR \cite{Aldrovandi_Pereira_2013} Lagrangians. It is quite permissible because both of them are scalar densities. However, only the conserved quantities related to the TEGR Lagrangian, that is Lorentz covariant, are free of the aforementioned problems, whereas the ones related to the Moller Lagrangian do not.

Advantages of the covariant TEGR Lagrangian and related conserved quantities appear due to incorporating the aforementioned inertial spin connection. However, the last is not determined by the construction. To close this gap, analyzing the structure of the new conserved quantities, we introduce a principle to determine it at the end of section 4. A necessity to determine   inertial spin connections appears in $f(T)$ theories as well. Thus, in \cite{f(T)5} one requires a consistence of field equations together with using symmetries of concrete solutions of  $f(T)$ theory. It is surprisingly, but these two quite different principles lead to the same results in cases of some concrete solutions, see sections \ref{mass} and \ref{observer}.

In section \ref{Comparison}, we compare incorporation of the inertial spin connection in the covariant TEGR with incorporation of background spacetime structures into a metric presentation of GR to covariantize Einstein's pseudotensor and Freud's superpotential.

In sections \ref{mass} and \ref{observer}, we apply new expressions in the covariant TEGR and our principle for determining the inertial spin connection to calculate the mass of the Schwarzschild black hole and densities of conserved quantities for a freely falling observer both in the Friedmann-Lema\^itre-Robertson-Walker (FLRW) universe and in (anti-)de Sitter ((A)dS) space. We compare inertial spin connections obtained on the basis of our principle in covariant TEGR with those obtained in \cite{f(T)5} with using symmetries in the Schwarzschild and the FLRW solutions in $f(T)$ theories.

In section \ref{Discussion}, we discuss new results, compare them with previous ones and give perspectives of their development.

In the study we are based of the variational principle and applying the Noether theorem. It is the standard and very economical way in deriving conservation laws and related conserved quantities in Lagrangian based theory. It gives a possibility to present an appropriate interpretation of results as well. In addition to our main study, in Appendix A, we demonstrate that our theoretical result, conservation laws (\ref{DiffCL_1}) and (\ref{DiffCL_2}), can be obtained {\em directly} starting from the field equations without applying the variational principle and Noether's theorem. By this, we independently check that our results are correct, and this helps us to understand them deeper.

\section{Preliminaries}
\setcounter{equation}{0}
\m{Preliminaries}

\subsection{Definitions and notations}

In the paper, definitions and notations mostly correspond to the ones in the book \cite{Aldrovandi_Pereira_2013}. Now we derive such of them which are necessary and important for our study. To represent an arbitrary metric theory (including GR) in a form of a tetrad theory one usually uses the relation
\be
g_{\mu\nu} = \eta_{ab}h^a{}_\mu h^b{}_\nu\,,
\m{g_hh_GR}
\ee
where $h^a{}_\mu$ are 16 tetrad components of  a new dynamical field instead of the metric components $g_{\mu\nu}$, and $\eta_{ab}$ is the Minkowski metric.

Here, we consider only GR and its tetrad representations. Following \cite{Aldrovandi_Pereira_2013}, we denote quantities related to GR itself by `overcircles'. Thus, Christoffel symbols (Levi-Civita zero-torsion connection) are denoted as usual,
\be
\cG{}^{\alf}{}_{\mu\nu} = \half g^{\alf\rho}\l(g_{\mu\rho,\nu} + g_{\nu\rho,\mu} - g_{\mu\nu,\rho} \r) \,.
\m{cG}
\ee
The curvature tensor associated with the connection (\ref{cG}) is denoted as $\cR{}^\alf{}_{\beta\mu\nu}$; the related Ricci tensor and curvature scalar are $\cR_{\mu\nu}$ and $\cR$, respectively. Covariant derivatives associated with (\ref{cG}) are denoted as $\cN_\mu$; thus, for example, applying it to contra-variant vector $u^\alf$, one derives $\cN_\mu u^\alf = \di_\mu u^\alf + \cG{}^{\alf}{}_{\mu\rho} u^\rho$.

The Levi-Civita {\em spin connection} is denoted as usual \cite{Landau_Lifshitz_1975}:
\be
\cA{}^i{}_{j\mu} = -h_j{}^\nu \cN_\mu h^i{}_\nu\,.
\m{circ_A}
\ee
For the sake of definiteness, below we call it the {\em GR spin connection}. The curvature tensor associated with the connection (\ref{circ_A}) is denoted as $\cR{}^a{}_{b\mu\nu}$.
One of the important rules is a conversion of the tetrad indices to spacetime ones and inversely with the use of tetrad vectors. Thus, applying this rule one can prove that, for example, $\cR{}^a{}_{b\mu\nu} = h^a{}_\alf h^\beta{}_b\cR{}^\alf{}_{\beta\mu\nu}$.

Return to the equality (\ref{g_hh_GR}), from where it follows $~\det{h^a{}_\mu} \equiv h=\sqrt{-g}$; $g\equiv\det{g}_{\mu\nu}$. If quantities $A^\alf$ and $B^{\alf\beta}$ are vector and antisymmetric tensor, respectively, then quantities ${\hat A}^\alf = hA^\alf= \sqrt{-g}A^\alf$ and ${\hat B}^{\alf\beta}= hB^{\alf\beta} = \sqrt{-g}B^{\alf\beta}$ are related densities of the mathematical weight $+1$. In our study, the equalities for divergences \cite{Landau_Lifshitz_1975},
\be
\di_\alf{\hat A}^\alf = \cN_\alf{\hat A}^\alf, \qquad \di_\beta{\hat B}^{\alf\beta} = \cN_\beta{\hat B}^{\alf\beta}\,,
\m{A_B}
\ee
are quite important. Indeed, to transform volume integration to surface one we use the Gauss theorem that is applied in the case of a partial divergence. The equalities, like (\ref{A_B}), give us the assurance that a covariance of expressions is preserved under a partial differentiation.

Here, considering covariant presentation of TEGR, we follow \cite{Aldrovandi_Pereira_2013} and denote quantities in covariant TEGR by `overdots'. It is important to introduce the Weitzenb\"ock spin connection $\sA{}^a {}_{b \mu}$; for a definiteness below we call it the {\em inertial spin connection}. In a so-called inertial frame one has $\sA{}^a {}_{b \mu} = 0$, see Chapter 2 in \cite{Aldrovandi_Pereira_2013}. From zeroth value, after a local Lorentz rotation the inertial spin connection becomes
\be
{\sA}{}^a{}_{c\mu} = \Lambda^a{}_b(x)\di_\mu \Lambda_c{}^b(x)
\m{iLconnect}
\ee
with a corresponding local Lorentz transformation matrix $\Lambda^b{}_d(x)$. Taking into account (\ref{iLconnect}), one constructs covariant with respect to local Lorentz rotations (Lorentz covariant) derivatives, $\sD_\sigma$. Thus, for example, applied to a tetrad co-vector, $V_a$, it is defined as
\be
\sD_\sigma V_a = \di_\sigma V_a - \sA{}^c{}_{a\sigma} V_c\,.
\m{sD}
\ee

The Weitzenb\"ock connection, $\sG{}^{\alf}{}_{\mu\nu}$, associated with the Weitzenb\"ock spin connection, ${\sA}{}^a{}_{c\mu}$, is defined as
\be
\sG{}^{\alf}{}_{\mu\nu} = h_a{}^\alf \l(\di_\nu h^a{}_\mu +  \sA{}^a{}_{c\nu} h^c{}_\mu \r) = h_a{}^\alf \sD_\nu h^a{}_\mu \,.
\m{sG}
\ee
The curvature tensor associated with the connection (\ref{sG}) is equal to zero,  $\sR{}^\alf{}_{\beta\mu\nu}\equiv 0$, identically; whereas the torsion tensor is nonzero, $\sT{}^{\alf}{}_{\mu\nu} = \sG{}^{\alf}{}_{\nu\mu}-\sG{}^{\alf}{}_{\mu\nu} \neq 0$, if the the gravity effects exist. Keeping in mind $\sR{}^\alf{}_{\beta\mu\nu}= 0$ and definition (\ref{sD}), one easily finds
\be
(\sD_\rho\sD_\sigma - \sD_\sigma\sD_\rho)V_a = 0\,.
\m{ssD}
\ee
We stress that the introduction of the inertial spin connection ${\sA}{}^a{}_{c\mu}$ (Lorentz not covariant quantity) allows us to construct Lorentz covariant quantities remarked by `overdots', like $\sL$, $\sT{}^{\alf}{}_{\mu\nu}$, ${\sK}{}^a{}_{c\mu}$, ${\sS}{}_a{}^{\rho\sig}$, and others, defined below in detail.


\subsection{Lagrangians in tetrad presentations of GR}
\m{Lagrangians}

It is well known that Einstein paid an essential attention to tetrad presentation of GR (teleparallel presentation), particularly, to unite electromagnetic interaction with gravity \cite{Einstein_1,Einstein_2}.  To represent GR in the teleparallel form we use the definition (\ref{g_hh_GR}) as well. Then the Hilbert Lagrangian is rewritten in the tetrad form \cite{Landau_Lifshitz_1975}:
\be
\lag_H = -\frac{h}{2\k}{\cR} = -\frac{h}{2\k} h_i{}^\mu h^{j\nu} \l( \di_\mu \cA{}^i{}_{j\nu} - \di_\nu \cA{}^i{}_{j\mu} + \cA{}^i{}_{k\mu}\cA{}^k{}_{j\nu} - \cA{}^i{}_{k\nu}\cA{}^k{}_{j\mu}\r)\,,
\m{Hilbert}
\ee

Lagrangian (\ref{Hilbert}) is of the second derivatives of $h^a{}_\mu$. By known reasons, it is more preferable to have Lagrangian of the first derivatives only. In this case: 1) Variation of the action is carried out under the Dirichlet boundary conditions when only fields, not their derivatives, are fixed at the boundary (on properties and advantages of the Dirichlet boundary conditions in GR see, for example, in \cite{JS}). 2) Energy-momentum has the first derivatives only as well (a related discussion see, for example, in \cite{KBL_1997,Babak_Grichshuk_2000}). It is important, for example, when the initial data problem is studied because initial fields and initial velocities define the energy-momentum at the initial hypersurface in an unambiguous way. At last, 3) the Lagrangian of the first derivatives can be used as a basis for constructing modified theories under a requirement to have field equations of the second order derivatives, not more (in the framework of the teleparallel approach, see, for example, review \cite{REV_2018}). Considering the Lagrangian (\ref{Hilbert}), one easily derives a related Lagrangian of the first derivatives:
\be
\lag_M = \lag_H - \di_\mu \hat m^\mu = \lag_H - \cN_\mu \hat m^\mu\,; \qquad \hat m^\mu = \frac{h}{\k}  h^{j\mu}h_i{}^\nu\cA{}^i{}_{j\nu}\,,
\m{Hilbert_div}
\ee
again, `hat` means that a quantity is a density of the weight $+1$. A replacement of the partial derivative with the covariant one is possible due to the equalities of the type (\ref{A_B}). Explicitly Lagrangian (\ref{Hilbert_div}) is rewritten as
\be
\lag_M = \frac{h}{2\k}h_i{}^\mu h^{j\nu}\l(\cA{}^i{}_{k\mu}\cA{}^k{}_{j\nu} - \cA{}^i{}_{k\nu}\cA{}^k{}_{j\mu}\r)\,
\m{Moller}
\ee
that is the well known Moller Lagrangian \cite{Moller_1961}.

Let us outline briefly Lagrangians (\ref{Hilbert}) and (\ref{Moller}). The Hilbert Lagrangian is a scalar density of the weight +1. Besides, it is Lorentz invariant, because it depends on the metric (\ref{g_hh_GR}) and its derivatives only, which are Lorentz invariant. Concerning the Moller Lagrangian (\ref{Moller}), it becomes not invariant with respect to local Lorentz rotations, whereas it continues to be a scalar density of the weight +1. Basing on the Lagrangian (\ref{Moller}), Moller has suggested the gravitational energy-momentum complex and related superpotential, which are covariant with respect to coordinate transformations. However, they, like the Lagrangian (\ref{Moller}), are not covariant with respect to local Lorentz rotations, for a more detail see subsection \ref{Moller_CL}.

To conserve the advantage of (\ref{Moller}) with a spacetime covariance and with the first derivatives only, but to restore the Lorentz invariance, one has to turn to the covariant TERG, see the book \cite{Aldrovandi_Pereira_2013}. We pay the main attention to the formalism developed in \cite{Aldrovandi_Pereira_2013} and do not touch principles of constructing covariant TEGR, like principles for constructing a gauge theory for the translation group; for discussing a philosophy of covariant TERG we again send a reader to the book \cite{Aldrovandi_Pereira_2013}. The basic formulae are as follow. Lorentz rotation for the tetrad components has the form:
\begin{equation}\label{lroth}
h'^a {}_{\mu} = \Lambda^a {}_b (x) h^b {}_\mu\,.
\end{equation}
Analogous transformations are applied to arbitrary tensors with the tetrad indices. Under a Lorentz rotation an arbitrary spin connection $A^a {}_{b \mu}$ is transformed as
\begin{equation}\label{lrotA}
A'^a {}_{b \mu} = \Lambda^a {}_c (x) A^c {}_{d \mu} \Lambda^d {}_b (x) + \Lambda^a {}_c(x) \partial_{\mu} \Lambda_b {}^c (x)\,.
\end{equation}

The GR spin connection (\ref{circ_A}) describes both the gravitational effects and the pure inertial effects. Therefore $\cA{}^a {}_{b \mu}$ cannot be equalized to zero totally by local rotations. On the other hand, the purely inertial spin connection $\sA{}^a {}_{b \mu}$ can be suppressed totally by a transformation, like (\ref{iLconnect}). Both the spin connection ${\sA}{}^a{}_{c\mu}$ and the spin connection ${\cA}{}^a{}_{c\mu}$ are not Lorentz covariant, see (\ref{lrotA}), but their difference
\be
{\sK}{}^a{}_{c\mu} = {\sA}{}^a{}_{c\mu} - {\cA}{}^a{}_{c\mu}\,
\m{contortion}
\ee
is Lorentz covariant. It is called the contortion tensor and vanishes if the gravitational effects are absent. In terms of ${\sK}{}^a{}_{c\mu}$ the Hilbert Lagrangian is rewritten in evidently Lorentz covariant form:
\be
\lag_H = -\frac{h}{2\k}{\cR} = \frac{h}{2\k} h_i{}^\mu h^{j\nu} \l( \sD_\mu \sK{}^i{}_{j\nu} - \sD_\nu \sK{}^i{}_{j\mu} + \sK{}^i{}_{k\mu}\sK{}^k{}_{j\nu} - \sK{}^i{}_{k\nu}\sK{}^k{}_{j\mu}\r)\,,
\m{Hilbert+}
\ee
where $\sD_\sigma$ is the Lorentz covariant differential operator; in (\ref{sD}), it is defined  as acting on a tetrad co-vector.

Again, to have Lagrangian of the first derivatives only one has to select a divergence,
\be
\sL = \lag_H - \di_\mu \hat t^\mu = \lag_H - \cN_\mu \hat t^\mu\,; \qquad \hat t^\mu = \frac{h}{\k}\sK{}^{\mu\nu}{}_{\nu}  \,.
\m{Hilbert_div+}
\ee
As a result, the Lagrangian (\ref{Hilbert_div+}) transforms to the explicit form:
\be
\sL = \frac{h}{2\kappa} \l({\sK}{}^\rho{}_{\mu\nu} {\sK}_\rho{}^{\nu\mu} - {\sK}{}^\nu{}_{\rho\nu} {\sK}{}^{\mu\rho}{}_\mu \r) \,.
\m{Hilbert_div+K}
\ee
Recall again that tetrad indices and spacetime indices are replaced by contracting with components of tetrad \cite{Aldrovandi_Pereira_2013}.
Frequently it is more convenient to use torsion tensor instead of contortion one:
\be
{\sT}{}^\rho{}_{\mu\nu} = {\sK}{}^\rho{}_{\mu\nu} - {\sK}{}^\rho{}_{\nu\mu}, \qquad 2{\sK}{}^\rho{}_{\mu\nu} = {\sT}_\mu{}^\rho{}_{\nu} + {\sT}_\nu{}^\rho{}_{\mu} - {\sT}{}^\rho{}_{\mu\nu}
\m{contor}
\ee
that is both spacetime and Lorentz covariant the same as the contortion tensor.
The explicit expression for the torsion tensor is
\be
{\sT}{}^a{}_{\mu\nu} = \di_\mu h^a{}_\nu - \di_\nu h^a{}_\mu + {\sA}{}^a{}_{c\mu}h^c{}_\nu - {\sA}{}^a{}_{c\nu}h^c{}_\mu\, .
\m{tor}
\ee
This expression due to the antisymmetry in $\mu$ and $\nu$ can be rewritten in evidently tensorial form:
\be
{\sT}{}^\rho{}_{\mu\nu} = h_a{}^\rho\l(\cN_\mu h^a{}_\nu - \cN_\nu h^a{}_\mu + {\sA}{}^a{}_{c\mu}h^c{}_\nu - {\sA}{}^a{}_{c\nu}h^c{}_\mu\r)\,.
\m{tor+}
\ee
Finally, Lagrangian (\ref{Hilbert_div+}) acquires the form:
\be
\sL = \frac{h}{2\kappa} \l(\frac{1}{4} {\sT}{}^\rho{}_{\mu\nu} {\sT}_\rho{}^{\mu\nu} + \frac{1}{2} {\sT}{}^\rho{}_{\mu\nu} {\sT}{}^{\nu\mu}{}_\rho - {\sT}{}^\rho{}_{\mu\rho} {\sT}{}^{\nu\mu}{}_\nu\r) \equiv - \frac{h}{2\kappa}T \,,
\m{lag}
\ee
that is the main form of the TEGR Lagrangian \cite{Aldrovandi_Pereira_2013}. Remark again that the Lagrangian (\ref{lag}), being of the first order derivatives only, is both spacetime covariant and Lorentz invariant, unlike the Lagrangian (\ref{Moller}).

At last, it is important to compare the Moller Lagrangian (\ref{Moller}) with the covariant TEGR one (\ref{Hilbert_div+K}), the same (\ref{lag}). Recall that under a related Lorentz rotation one can set ${\sA}{}^a{}_{b\mu} = 0$. Of course, under the same local Lorentz rotation the GR spin connection ${\cA}{}^a{}_{b\mu}$ and tetrad vectors $h^a{}_\mu$ are transformed respectively. Such a fixation, when ${\sA}{}^a{}_{b\mu} = 0$, is usually called the {\em Weitzenb\"ok gauge} in TEGR, and then one easily finds that the Lagrangian (\ref{Hilbert_div+K}) transforms to (\ref{Moller}). In spite of that there is a basic difference. Indeed, after generalizing the Weitzenb\"ok gauge one has ${\sA}{}^a{}_{b\mu} \neq 0$ and the covariant TEGR Lagrangian again becomes of the form (\ref{Hilbert_div+K}), whereas the Moller Lagrangian saves the form (\ref{Moller}). Thus, below we will distinguish the Moller presentation from the covariant TEGR presentation and inversely, and compare them.

\subsection{The field equations}
\m{FieldEquations}

For the sake of generality, let us include matter sources in both the cases of Lagrangians $\lag_M$ in (\ref{Moller}) and $\sL$ in (\ref{lag})  as follows,
\bea
\lag_{\scriptscriptstyle MTOT} &=& \lag_M + \lag_m\l(g_{\mu\nu}(h), \phi\r)\,,\m{lag_MTOT} \\
  \lag_{\scriptscriptstyle TEGR} &=& \sL + \lag_m\l(g_{\mu\nu}(h), \phi\r)\,,
 \m{lag_TEGR}
\eea
with generalized matter fields $\phi$; the tetrad field is included into $\lag_m$ over the metric (\ref{g_hh_GR}) only. Let us compare $\lag_M$ and $\sL$ using (\ref{Hilbert_div}) and (\ref{Hilbert_div+}):
\be
\sL = \lag_M - \di_\mu \l(\hat m^\mu - \hat t^\mu \r) = \lag_M + \frac{1}{\k}\di_\mu \l(h h_i{}^\mu h^{j\nu} \sA{}^i{}_{j\nu} \r)\,.
\m{sL_mL}
\ee
One can see that the inertial spin connection is included in the Lagrangian of the covariant TEGR only under the total divergence. Thus, one concludes that $\lag_M$ is Lorentz covariant, however, up to a total divergence only.

Let us discuss a necessity/unnecessity to vary the divergence in (\ref{sL_mL}). In the framework of metric GR, in attempts to define conserved quantities for an {\em asymptotically flat spacetime} many studies have been provided. In this relation, a big opinion has been paid to {\em Hamiltonian formulation of GR}, the more important papers are \cite{Regge_Teitelboim_1974,Faddeev_1982} and references there in. Then, indeed, under variation one has to take into account a divergence in the Hamiltonian (surface term in Hamiltonian action) and fall-off conditions for metric components. Unlike this, we apply the Lagrangian formalism only, and we do not restrict ourselves by asymptotically flat models. Thus, we follow the prescriptions in many textbooks (see, for example, \cite{Landau_Lifshitz_1975}) when variation of a total divergence with respect to field variables (here $h^a{}_\rho$) gives zero.

Concerning variation of the total divergence in (\ref{sL_mL}) with respect to $\sA{}^i{}_{j\nu}$ our position is that it has to give zero as well. Indeed, in whole the inertial spin connection does not participate in field equations (both in covariant TEGR and in Moller's TEGR); besides, $\sA{}^i{}_{j\nu}$ has no an influence on construction of concrete solutions. Thus, components of $\sA{}^i{}_{j\nu}$ are undetermined functions in TEGR itself. Then, the variation of the divergence in (\ref{sL_mL}) with respect to $\sA{}^i{}_{j\nu}$ has to give $~0=0~$ that is quite acceptable. In the framework of TEGR, many authors follow this assertion as well, see, for example, \cite{Krssak_2015}; in review \cite{REV_2018} authors explain this situation as {\em ``...and hence there are no extra field equations that would determine it {\em [spin connection]}. This is consistent with our interpretation of the spin connection as representing inertial effects only, and hence should not have their own dynamics governed by extra field equations.''}

Keeping in mind all the above, we state the following. In our study of TEGR, the field $h^a{}_\rho$ is an unique dynamic gravitational field. Indeed, the field gravitational equations are obtained by varying with respect to $h^a{}_\rho$ only; solutions to these equations are constructed with using $h^a{}_\rho$ only as well (adding matter variables if necessary). Then, in the framework of the formal consideration, the inertial spin connection ${\sA}{}^i{}_{j\mu}$ defined in (\ref{iLconnect}) can be classified as a passive (auxiliary) field.\footnote{On interpretation of the inertial spin connection as representing inertial effects see book \cite{Aldrovandi_Pereira_2013} and review \cite{REV_2018} that does not contradict to our formal interpretation.} Making the use of ${\sA}{}^i{}_{j\mu}$ we derive {\em explicitly} Lorentz covariant field equations (see below (\ref{EM}));  with the use of ${\sA}{}^i{}_{j\mu}$ we construct Lorentz invariant conserved quantities in covariant TEGR (see below (\ref{DiffCL_1}) and (\ref{DiffCL_2})).

To provide variation of  (\ref{lag_MTOT}) and  (\ref{lag_TEGR}) with respect to $h^a{}_\rho$ one needs to derive
\bea
\delta h_b{}^\rho &=& - h_a{}^\rho h_b{}^\mu \delta h^a{}_\mu\,, \nonumber \\
\delta g_{\mu\nu} &=&  2h_{a(\mu}\delta h^a{}_{\nu)}\,, \nonumber \\
\delta g^{\mu\nu} &=& - 2g^{\rho(\mu} h_a{}^{\nu)}\delta h^a{}_{\rho}\,.\nonumber
\eea
Then one easily obtains the gravitational field equations:
\bea
\frac{\delta \lag_M}{\delta h^a{}_\rho} &\equiv & \frac{\di \lag_M}{\di h^a{}_\rho} - \di_\sigma \l( \frac{\di \lag_M}{\di h^a{}_{\rho,\sigma}}\r) =
h\Theta_a{}^\rho \,,
\m{MollerEM}\\
\frac{\delta \sL}{\delta h^a{}_\rho} &\equiv & \frac{\di \sL}{\di h^a{}_\rho} - \di_\sigma \l( \frac{\di \sL}{\di h^a{}_{\rho,\sigma}}\r) =
h\Theta_a{}^\rho \,,
\m{EM}
\eea
where the matter energy-momentum tensor is defined as
\be
\Theta_a{}^\rho = -\frac{1}{h}\frac{\delta \lag_m}{\delta h^a{}_\rho}\,.
\m{matterEM}
\ee
Recalling the discussion of the divergence in (\ref{sL_mL}), we conclude that the left hand sides of equations of motion (\ref{MollerEM}) and (\ref{EM}) are the same, only  (\ref{EM}) is presented in the {\em explicitly} Lorentz covariant form, unlike (\ref{MollerEM}). Thus, on the one hand, operators at the left hand sides of both these equations do not depend on the inertial spin connection ${\sA}{}^i{}_{j\mu}$ in whole; on the other hand, both these operators are Lorentz covariant anyway. It is not surprisingly because $\lag_M$ is Lorentz invariant up to a divergence as it has been noted above. Of course, both (\ref{MollerEM}) and (\ref{EM}) are equivalent to the Einstein equations in the standard metric formulation.

 \subsection{Problems of conservation laws in TEGR}

As it has been noted in Introduction, a construction of conserved quantities in TEGR has problems. Concerning the Moller presentation and analyzing the equations (\ref{MollerEM}), we send a reader to the Moller paper \cite{Moller_1961}. Thus, let us turn to the covariant TEGR and consider (\ref{EM}) in a more detail. Let us introduce new notations as
\bea
\frac{\di \sL}{\di h^a{}_\rho}& \equiv &- h \ssJ_a{}^\rho\,,\m{J_bullet} \\
\frac{\di \sL}{\di h^a{}_{\rho,\sigma}} &\equiv &- \frac{h}{\kappa} \sS_a{}^{\rho\sigma}\,.\m{S_bullet}
\eea
 Providing the direct calculations, one obtains
\bea
\ssJ_a{}^\rho&=& - \frac{h_a{}^\rho}{h}\sL + \frac{1}{\kappa} h_a{}^\pi\sS_c{}^{\beta\rho}\sT{}^c{}_{\beta\pi} +  \frac{1}{\kappa} \sA{}^c{}_{a\beta}\sS_c{}^{\rho\beta} \,,     \m{J}\\
 \sS_a{}^{\rho\sigma} &=& \l(\half  {\sT}_\beta{}^{\rho\sigma} -  {\sT}{}^{[\rho\sigma]}{}_\beta - 2{\sT}{}^{\alpha[\rho}{}_\alpha \delta^{\sigma]}_\beta  \r)h_a{}^\beta\,, \m{S}
 \eea
 where the last expression is antisymmetric in the upper indices.
Thus, (\ref{EM}) can be rewritten in the form:
 \be
 \kappa h \l(\ssJ_a{}^\rho + \Theta_a{}^\rho\r)= \di_\sigma \l(h \sS_a{}^{\rho\sigma}\r)\,
 \m{EM+}
 \ee
that is really presents the conservation law. Indeed, its divergence gives
 \be
 \di_\rho \l(h \ssJ_a{}^\rho + h\Theta_a{}^\rho\r) = 0\,.
 \m{CL+}
 \ee

Expressions in (\ref{EM+}) and (\ref{CL+}), in spite of partial derivatives, are coordinate covariant. Indeed, it is easily to represent (\ref{EM}), (\ref{EM+}) and (\ref{CL+}) in evidently spacetime covariant form:
\bea
&&\frac{\di \sL}{\di h^a{}_\rho} - \cN_\sigma \l( \frac{\di \sL}{\di h^a{}_{\rho,\sigma}}\r) = h\Theta_a{}^\rho\,, \m{EM_c}\\
&&\kappa h \l(\ssJ_a{}^\rho + \Theta_a{}^\rho\r)= \cN_\sigma \l(h \sS_a{}^{\rho\sigma}\r)\,, \m{EM_c+}\\
&&\cN_\rho \l(h \ssJ_a{}^\rho+ h\Theta_a{}^\rho\r) = 0\,.
 \m{CL_c}
 \eea
 A replacement of partial derivatives in (\ref{EM}), (\ref{EM+}) and (\ref{CL+}) with the covariant ones is possible due to the equalities of the type (\ref{A_B}). Indeed, partial derivatives of vector densities and antisymmetric tensor densities can be exchanged by covariant ones. Integration of (\ref{EM+}) and (\ref{CL+}) by the standard methods \cite{Landau_Lifshitz_1975,Misner_Thorn_Wheeler_1973} gives integral conserved quantities on hypersurfaces $x^0 = {\rm const} := \Sigma$:
 \be
 {\cal P}_a = \int_\Sigma dx^3 h\l(\ssJ_a{}^0 + \Theta_a{}^0\r) = \frac{1}{\k}\oint_{\di\Sigma} ds_i \l(h \sS_a{}^{0i}\r)\,.
 \m{ICQ}
 \ee
This construction was possible namely due to the spacetime covariance.

Note that the quantity (\ref{J}), being a spacetime vector, is not Lorentz covariant. Keeping in mind (\ref{tor}) - (\ref{lag}), one easily recognizes that the left hand side of (\ref{S_bullet}) is both Lorentz and coordinate covariant. Thus, the quantity (\ref{S}) is antisymmetric spacetime tensor and it is Lorentz covariant. How can one explain the Lorentz non-covariance of the left hand side of (\ref{EM+})? It is because that the partial differentiation in the right hand side acts onto expression with a tetrad index. The authors of the book \cite{Aldrovandi_Pereira_2013} rewrite (\ref{J}) as
\bea
\ssJ_a{}^\rho&=& \sJ{}_a{}^\rho +  \frac{1}{\kappa} \sA{}^c{}_{a\beta}\sS_c{}^{\rho\beta} \,,     \m{J+}\\
\sJ{}_a{}^\rho&\equiv & - \frac{h_a{}^\rho}{h}\sL + \frac{1}{\kappa} h_a{}^\pi\sS_c{}^{\beta\rho}\sT{}^c{}_{\beta\pi}\,.
\m{J*}
\eea
Then we rewrite the equation (\ref{EM+}) in terms of Lorentz covariant expressions:
\be
 \kappa h \l(\sJ_a{}^\rho + \Theta_a{}^\rho\r)= \di_\sigma \l(h \sS_a{}^{\rho\sigma}\r) -\sA{}^c{}_{a\beta}\l(h\sS_c{}^{\rho\beta}\r) = \sD_\sigma \l(h \sS_a{}^{\rho\sigma}\r)\,,
 \m{EM_Lc}
 \ee
where the definition (\ref{sD}) has been used.  Because the quantity under the divergence is antisymmetric in upper indices, one finds that, due to the equality of the type (\ref{ssD}), the divergence $\sD_\rho$ of (\ref{EM_Lc}) makes the right hand side of (\ref{EM_Lc}) zeroth one. This leads to the conservation law:
 \be
 \sD_\rho\l( h \sJ{}_a{}^\rho + h\Theta_a{}^\rho\r)  = 0\,.
 \m{J_Lc}
 \ee
However, now we cannot integrate both (\ref{EM_Lc}) and (\ref{J_Lc}) following (\ref{ICQ}) with the goal to obtain conserved charges. In subsection \ref{TEGR_CL}, we provide a construction of conserved quantities in the covariant TEGR, which are simultaneously spacetime covariant,  Lorentz covariant and allows us to construct well defined charges.

\section{The Noether theorem in a covariant field theory}
\setcounter{equation}{0}
\m{Noether}

Here, to analyze spacetime covariant Lagrangians (\ref{Moller}) and (\ref{lag}) we present Noether's identities in an arbitrary covariant theory with an action
\be
S = \int dx^4 \lag(\psi^A;\psi^A{}_{,\alpha})\,.
\m{Sarbitrar}
\ee
The notation $\psi^A$ means an arbitrary tensor density or a set of such densities, where $A$ is a collective index. We derive conserved quantities and related conservation laws that follow from diffeomorphism invariance, see \cite{Petrov_KLT_2017,Mitskevich_1969,Petrov_Lompay_2013}. We consider the variations of $\psi^A$ in the form of the Lie derivative
 \be
  \delta \psi^A= {\pounds}_\xi \psi^A =-\xi^\alf \di_\alf \psi^A +
 {\l. \psi^A \r|}^\alf_\beta \di_\alf \xi^\beta\,,
\m{a-d4}
 \ee
 note that we use the opposite signs with respect to the standard ones.

 The description of  the notations in (\ref{a-d4}), like ${\l. \psi^A \r|}^\alf_\beta$, and their algebra can be found in Appendix 3.1 of the book \cite{Petrov_KLT_2017}. Here, it is enough to know concrete transformation properties of  $\psi^A$. Let us assume that the set $\psi^A$ is presented by the vector $\phi^\sigma$ only. Then the Lie derivative in the formula (\ref{a-d4}) is defined by ${\l. \phi^\sigma \r|}^\alf_\beta = \delta^\sigma_\beta \phi^\alpha$. Analogously for the tensor $\psi^A = \pi_{\mu\nu}$ one has ${\l. \pi_{\mu\nu} \r|}^\alf_\beta = - \delta^\alf_\mu \pi_{\beta\nu}- \delta^\alf_\nu \pi_{\mu\beta}$, etc. Because, the Lie operator (\ref{a-d4}) is linear, the notation ${\l. \psi^A \r|}^\alf_\beta$ is generalized easily to an arbitrary set of tensor densities, for example, like $\psi^A = \l\{\phi^\sigma,\ldots ,\pi_{\mu\nu}\r\}$. The notation ${\l. \psi^A \r|}^\alf_\beta$ is quite compact and significantly simplifies calculations, see Appendix 3.1 of the book \cite{Petrov_KLT_2017}.

Here, under the general consideration, we do not concretize fields in the set $\psi^A$, they can be both dynamical ones (field variables of the theory) and external auxiliary or passive fields. However it is assumed that all the fields presenting set $\psi^A$ are geometrical objects. Then all of them must be taken into account under the Lie procedure, and for all of them variations, like (\ref{a-d4}), must appear. It is because the Lie procedure acts onto {\em all} the geometrical objects defined on the spacetime manifold. Especially we remark that we do not consider here the explicit dependence of the Lagrangian on spacetime coordinates, like $\lag =\lag(\psi^A; \psi^A_{,\alf}; x^\alf)$.

Because $\lag$ is a scalar density of the weight +1 the diffeomorphism invariance leads to the main Noether's identity:
\be
\Lix \lag \equiv - \di_\alf\l(\xi^\alf \lag \r)\,.
\m{a-b27}
\ee
This identity after substituting (\ref{a-d4}) can be rewritten in the form:
 \be
\frac{\delta\lag}{\delta \psi^B}{\pounds}_\xi \psi^B + \di_\alf \l[{{\di \lag} \over {\di
\psi^{B}{}_{,\alf}}}{\pounds}_\xi \psi^B +
\xi^\alf {\lag}\r]\equiv 0\,.
 \m{a-d5}
 \ee
Substituting (\ref{a-d4}) into (\ref{a-d5}), one gets
 \be
-\l[\frac{\delta\lag}{\delta \psi^B} \psi^{B}{}_{,\alf} + \di_\beta \l(\frac{\delta\lag}{\delta \psi^B}
\l.\psi^{B}\r|^\beta_\alf\r)\r]\xi^\alf +\di_\alf \l[{\cal U}_\sig{}^\alf\xi^\sig + {\cal
M}_{\sig}{}^{\alf\tau}\di_\tau \xi^\sig \r] \equiv 0\,,
 \m{a-d7}
 \ee
 where the coefficients under the divergence are fully determined by the Lagrangian,
 \bea {\cal U}_\sig{}^\alf & \equiv & \lag \delta^\alf_\sig +
 {{\delta \lag} \over {\delta \psi^B}} \l.\psi^B\r|^\alf_\sig -
 {{\di \lag} \over {\di  \psi^{B}{}_{,\alf}}} \di_\sig \psi^{B}  \, , \m{a-d8}\\
 {\cal M}_\sig{}^{\alf\tau} & \equiv &
 {{\di \lag} \over {\di  \psi^{B}{}_{,\alf}}}
 \l.\psi^{B}\r|^\tau_\sig \, .
\m{a-d9}
 \eea

 Executing the operation of the partial derivative in the identity (\ref{a-d7}) and taking into account that the vector field $\xi^\sig$, and all its partial derivatives are independent and arbitrary at each point of spacetime manifold,
we come to the conclusion that all coefficients coupled with $\xi^\sig$, and its partial derivatives must be separately equal to zero.
It yields the system of identities:
 \bea
 &{}& \di_\alf  {\cal U}_\sig{}^\alf
  \equiv \frac{\delta\lag}{\delta \psi^B} \psi^{B}{}_{,\sig} + \di_\beta
\l(\frac{\delta\lag}{\delta \psi^B}
\l.\psi^{B}\r|^\beta_\sig\r), \m{a-d11}\\
&{}&    {\cal U}_\sig{}^\alf + \di_\lam {\cal M}_{\sig}{}^{\lam \alf} \equiv 0,
 \m{a-d12}\\ &{}&
 {\cal M}_{\sig}{}^{(\alf\beta)} \equiv 0. \m{a-d13}
 \eea
Thus, the coefficient (\ref{a-d9}) is antisymmetric in two last indices.

 The system (\ref{a-d11}) - (\ref{a-d13}) was engineered by Klein, see general and detail discussion in \cite{Lompay_Petrov_2013,Lompay_Petrov_2013A}.
Therefore, we refer to this system as the {\em Klein
identities}. After differentiating (\ref{a-d12}) and using
(\ref{a-d13}) one obtains that
$\di_\alf  {\cal U}_\sig{}^\alf \equiv 0$. This means that the right
hand side of (\ref{a-d11}) must be equal to zero identically as well,
 \be
 \frac{\delta\lag}{\delta \psi^B} \psi^{B}{}_{,\alf} + \di_\beta
\l(\frac{\delta\lag}{\delta \psi^B} \l.\psi^{B}\r|^\beta_\alf\r) \equiv 0\;.
 \m{a-d15}
 \ee
 It is just the claim of the second Noether's theorem \cite{Konopleva_Popov_1981}, besides it is a generalization of the Bianchi identity. Taking into account the historic development, we call the system (\ref{a-d11}) - (\ref{a-d15}) as the {\em Klein-Noether identities}\index{Klein-Noether system of identities}.

 The identity (\ref{a-d15}) suggests that instead of (\ref{a-d7}) one can use independently (\ref{a-d15}) and
  \be
  \di_\alf \l[{\cal U}_\sig{}^\alf\xi^\sig + {\cal M}_{\sig}{}^{\alf\tau}\di_\tau \xi^\sig \r] \equiv 0\,.
 \m{a-d16}
 \ee
The vector density entering under the divergence is classified as the current
 \be
{\cal I}^\alf(\xi) \equiv -\l[{\cal U}_\sig{}^\alf\xi^\sig + {\cal M}_{\sig}{}^{\alf\tau}\di_\tau \xi^\sig \r]\,.
 \m{a-d17}
 \ee
 The minus sign is chosen for making a correspondence with the usual minus sign in front of the gravitational
(metric) action, see, for example, (\ref{Hilbert}). Thus, the identity (\ref{a-d16}) is rewritten as
  \be
\di_\alf{\cal I}^\alf(\xi)\equiv 0\,.
 \m{a-d16_a}
 \ee
Because it is the identity, the current\index{current} has to be expressed through a divergence of a quantity (superpotential), ${\cal I}^\alf(\xi) \equiv \di_{\beta}{\cal I}^{\alf\beta}(\xi)$, a double divergence of which has to be equal to zero identically:  $\di_{\alf\beta}{\cal I}^{\alf\beta}(\xi)\equiv 0$. Let us show this.

Substituting (\ref{a-d12}) into (\ref{a-d17}), one obtains
 \be
{\cal I}^\alf(\xi) \equiv \di_\beta\l( {\cal M}_\sig{}^{\beta\alf}\xi^\sig  \r)\,.
 \m{a-d19}
 \ee
Because ${\cal M}_\sig{}^{\alf\beta}$ is antisymmetric in $\alf$ and $\beta$ the identity (\ref{a-d19}) can be rewritten in the form:
 \be
 {\cal I}^\alf(\xi) \equiv \di_\beta  {\cal I}^{\alf\beta}(\xi)
 \m{a-d21}
 \ee
 where the superpotential is
 \be
  {\cal I}^{\alf\beta}(\xi) = -{\cal M}_\sig{}^{\alf\beta}\xi^\sig  \,.
 \m{a-d22}
\ee
  One can see that the identity (\ref{a-d16_a}) for the current follows from the identity (\ref{a-d21}). The conservation law (\ref{a-d16_a}) permits to define a conserved quantity, ${\cal P}(\xi)$, on a hypersurface $\Sigma := t=constant$:
 \be  {\cal P}(\xi) = \int_\Sigma d^3x {\cal I}^{0}(\xi)\,.
 \m{a-d24_kops}
 \ee
By (\ref{a-d21}) it is effectively reduced to a surface integral:
 \be  {\cal P}(\xi) =  \oint_{\di\Sigma} ds_i {\cal I}^{0i}(\xi)\,
 \m{a-d24}
 \ee
that is called as a conserved charge.

\section{Conserved quantities}
\setcounter{equation}{0}
\m{CL}

In this section, basing on the results of previous section, we construct conserved quantities both in the tetrad GR in the Moller presentation and in the covariant TEGR presentation and compare them.

\subsection{The Moller conserved quantities}
\m{Moller_CL}

The Moller Lagrangian (\ref{Moller}), being spacetime covariant, is of the kind in (\ref{Sarbitrar}) considered above. The collective field $\psi^A$ is presented by a tetrad of covariant vectors: $\psi^A = \{h^a{}_\rho\}$, no other fields here.
Variations (\ref{a-d4}) are rewritten as
 \be
\delta h^a{}_\rho = {\pounds}_\xi h^a{}_\rho =-\xi^\alf  h^a{}_{\rho,\alf} -\xi^\alf{}_{,\rho}h^a{}_{\alf}\,.\m{a-d4_h+}
 \ee
Then, the coefficients  (\ref{a-d8}) and (\ref{a-d9}) related to the Moller Lagrangian (\ref{Moller}) are
 \bea
 {\MU}_\sig{}^\alf & \equiv &   \lag_M \delta^\alf_\sig +
 {{\delta \lag_M} \over {\delta h^a{}_{\rho}}} \l. h^a{}_{\rho}\r|^\alf_\sig - {{\di \lag_M} \over {\di  h^a{}_{\rho,\alf}}} h^a{}_{\rho,\sig}
\, , \m{a-d8_hAA}\\
{\MM}_\sig{}^{\alf\tau} &\equiv &
 {{\di \lag_M} \over {\di  h^a{}_{\rho,\alf}}} \l.h^a{}_{\rho}\r|^\tau_\sig \equiv
- {{\di \lag_M} \over {\di  h^a{}_{\tau,\alf}}} h^a{}_{\sig} \,.
\m{a-d9_M}
 \eea
Note that the quantity
  \be
 {{\di \lag_M} \over {\di  h^a{}_{\beta,\alf}}} = -\frac{h}{2\k}\cA{}^i{}_{j\mu}\l[h_a{}^\mu\l( h_i{}^\alf h^{j\beta} - h_i{}^\beta h^{j\alf}\r) -2
h_i{}^\mu\l(h_a{}^\alf h^{j\beta}- h_a{}^\beta h^{j\alf} \r) \r]\,
\m{super_M+}
\ee
in the expressions (\ref{a-d8_hAA}) and (\ref{a-d9_M}) is spacetime covariant, although it is not Lorentz covariant.

Finally, the current (\ref{a-d17}) related to the Moller Lagrangian acquires the form:
 \be
{\MJ}{}^\alf(\xi) \equiv \l[h\Mt_\sig{}^\alf + {{\delta \lag_M} \over {\delta h^a{}_{\alf}}} h^a{}_{\sigma}\r]\xi^\sigma + {\MM}_\sig{}^{\rho\alf}\cN_\rho\xi^\sig\,,
 \m{current_M}
 \ee
  where the Moller \cite{Moller_1961} gravitational energy-momentum tensor is defined as
 \be
 \Mt_\sig{}^\alf  \equiv  \frac{1}{h}\l[{{\di \lag_M} \over {\di  h^a{}_{\rho,\alf}}} \cN_\sig h^a{}_{\rho} - \lag_M \delta^\alf_\sig\r]
\, . \m{MollerTEI}
 \ee
By the general formulae (\ref{a-d8}) and (\ref{a-d17}), the current (\ref{current_M}) is presented with the partial derivatives $\di_\sig h^a{}_{\rho}$ and $\di_\rho\xi^\sig$ only. However, one can easily show that it is equivalent to the evidently covariant form (\ref{current_M}) with $\cN_\sig h^a{}_{\rho}$ and $\cN_\rho\xi^\sig$.

The current (\ref{current_M}) is conserved identically
  \be
\di_\alf {\MJ}{}^\alf(\xi) \equiv \cN_\alf {\MJ}{}^\alf(\xi) \equiv 0\,.
\m{CL_current_M}
\ee
It, as usual, is expressed identically through a divergence of the superpotential:
  \be
{\MJ}{}^\alf(\xi) \equiv \di_\alf {\MJ}{}^{\alf\beta}(\xi) \equiv \cN_\alf {\MJ}{}^{\alf\beta}(\xi)\,,
\m{CL_M}
\ee
where the superpotential, see (\ref{a-d22}), is
  \be
{\MJ}{}^{\alf\beta}(\xi) = -{\MM}_\sig{}^{\alf\beta}\xi^\sig \,.
\m{super_M}
\ee

All the derived above expressions are spacetime covariant, however they are evidently Lorentz non-covariant. Thus, first, we support the claim by Moller \cite{Moller_1961} that the energy-momentum (\ref{MollerTEI}) and the superpotential (in the Moller interpretation) (\ref{a-d9_M}) are not covariant with respect to local Lorentz rotations. Second, we make more wide assertion that both the current (\ref{current_M}) and the superpotential (\ref{super_M}) are not Lorentz covariant as well.


The above conservation laws (\ref{CL_current_M}) and (\ref{CL_M}) are identities only, they do not bring a physical content because up to now the field
equations have not been used. After using the equations (\ref{EM})  the current (\ref{current_M}) transforms to
 \be
{\MJ}{}^\alf(\xi) = h\l(\Mt_\sig{}^\alf  + {\Theta}_\sig{}^\alf \r) \xi^\sigma + {\MM}_\sig{}^{\rho\alf}\cN_\rho\xi^\sig\,.
 \m{current_M+}
 \ee
  As a result the identities (\ref{CL_current_M}) and (\ref{CL_M}) become physically sensible differential conservation laws:
  \bea
\di_\alf {\MJ}{}^\alf(\xi) &=& 0\,,
\m{MCL_1}\\
{\MJ}{}^\alf(\xi) &=& \di_\alf {\MJ}{}^{\alf\beta}(\xi)\,.
\m{MCL_2}
\eea
%

At last, let us make a remark. To improve the situation related to Lorentz non-covariance one could use the {\em reference} spin connection $\bar A{}^i{}_{j\mu}$ and  modify the  superpotential (\ref{super_M}) as
  \be
{\MJ}{}^{*\alf\beta}(\xi) = -\frac{h}{2\k}\l(\cA{}^i{}_{j\mu}- \bar A{}^i{}_{j\mu}\r)\l[\delta^\mu_\sigma\l( h_i{}^\alf h^{j\beta} - h_i{}^\beta h^{j\alf}\r) -2
h_i{}^\mu\l(\delta^\alf_\sig h^{j\beta}- \delta^\beta_\sig h^{j\alf} \r) \r]\xi^\sigma\,.
\m{super-bar}
\ee
This quantity is both spacetime covariant and Lorentz covariant. Then, first, because it has no tetrad indices its divergence is also Lorentz covariant; second, due to antisymmetry its double divergence disappears. The conservation law
  \be
{\MJ}{}^{*\alf}(\xi) \equiv \di_\beta {\MJ}{}^{*\alf\beta}(\xi) \equiv \cN_\beta {\MJ}{}^{*\alf\beta}(\xi)
\m{CL_M_bar}
\ee
allows us to define the current ${\MJ}{}^{*\alf}(\xi)$ as both spacetime covariant and Lorentz covariant.

\subsection{Conserved quantities in the covariant TEGR}
\m{TEGR_CL}

Lagrangian (\ref{lag}) is of the kind (\ref{Sarbitrar}) as well. Therefore, the technique of the section 3 can be directly applied in the covariant TEGR. Then the collective field $\psi^A$ is presented as a set of spacetime covariant vectors: $\psi^A = \{h^a{}_\rho;\sA{}^a{}_{b\mu}\}$. In spite of that inertial spin connection $\sA{}^a{}_{b\mu}$ is not a dynamic field one needs take it into account because diffeomorphisms act on all the geometrical objects under consideration.

Thus, variations (\ref{a-d4}) are rewritten as
 \bea
  \delta h^a{}_\rho&=& {\pounds}_\xi h^a{}_\rho =-\xi^\alf  h^a{}_{\rho,\alf} -\xi^\alf{}_{,\rho}h^a{}_{\alf}\,,\m{a-d4_h}\\
 \delta \sA{}^a{}_{b\mu}&=& {\pounds}_\xi \sA{}^a{}_{b\mu} =-\xi^\alf \sA{}^a{}_{b\mu,\alf} -\xi^\alf{}_{,\mu} \sA{}^a{}_{b\alf}\,.
\m{a-d4_A}
 \eea
Then, the coefficients (\ref{a-d8}) and (\ref{a-d9}) related to the covariant TEGR Lagrangian (\ref{lag}) become
 \bea
 {\sU}_\sig{}^\alf  &\equiv  &\sL \delta^\alf_\sig +  {{\delta \sL} \over {\delta h^a{}_{\rho}}} \l. h^a{}_{\rho}\r|^\alf_\sig  +
 {{\di \sL} \over {\di \sA^a{}_{b\mu}}} \l. \sA^a{}_{b\mu}\r|^\alf_\sig -
 {{\di \sL} \over {\di  h^a{}_{\rho,\alf}}} h^a{}_{\rho,\sig} \, , \m{a-d8_hA}\\
{\sM}_\sig{}^{\alf\tau}  &\equiv &
{{\di \sL} \over {\di  h^a{}_{\rho,\alf}}} \l.h^a{}_{\rho}\r|^\tau_\sig\,.
 \eea
Next, taking into account (\ref{lag}) and (\ref{tor}) one concludes
\be
{{\di \sL} \over {\di \sA^a{}_{b\mu}}} \l. \sA^a{}_{b\mu}\r|^\alf_\sig = -
 {{\di \sL} \over {\di  h^a{}_{\rho,\alf}}} \sA^a{}_{b\sigma}h^b{}_\rho\,.
 \m{sL}
\ee
Then, the current (\ref{a-d17})  acquires the form:
 \be
{\scJ}{}^\alf(\xi) \equiv \l[
 h\stheta_\sig{}^\alf + {{\delta \sL} \over {\delta h^a{}_{\alf}}} h^a{}_{\sigma}\r]\xi^\sigma + {{\di \sL} \over {\di  h^a{}_{\rho,\alf}}} h^a{}_{\sig}\cN_\rho\xi^\sig\,,
 \m{current_c}
 \ee
 where the quantity
 \be
 \stheta_\sig{}^\alf \equiv \frac{1}{h}
 \l[{{\di \sL} \over {\di  h^a{}_{\rho,\alf}}} \l( \cN_\sig h^a{}_{\rho} + \sA^a{}_{b\sigma}h^b{}_\rho\r) - \sL \delta^\alf_\sig \r]\,
 \m{E-M_h}
 \ee
can be interpreted as the energy-momentum tensor of the gravitational field in the covariant TEGR. Initially the current (\ref{current_c}) was presented with the partial derivatives $\di_\sig h^a{}_{\rho}$ and $\di_\rho\xi^\sig$ only. However, one can easily show that it is equivalent to the evidently covariant form (\ref{current_c}) with $\cN_\sig h^a{}_{\rho}$ and $\cN_\rho\xi^\sig$.

Due to the definitions (\ref{contortion}) and (\ref{S_bullet}) the current (\ref{current_c}) and the energy-momentum (\ref{E-M_h})
can be represented as
 \bea
{\scJ}{}^\alf(\xi) &\equiv &\l[ h\stheta_\sig{}^\alf
  + {{\delta \sL} \over {\delta h^a{}_{\alf}}} h^a{}_{\sigma}\r]\xi^\sigma + \frac{h}{\k} \sS_{\sig}{}^{\alf\rho}\cN_\rho\xi^\sig\,,
 \m{current_cL}\\
 \stheta_\sig{}^\alf &\equiv &\frac{1}{\k}\sS_{a}{}^{\alf\rho}{\sK}^a{}_{\sig\rho}- \frac{1}{h}\sL \delta^\alf_\sig \,,\m{E_M+}
 \eea
both of them are explicitly spacetime covariant and Lorentz covariant.

The current (\ref{current_cL}) is identically conserved, see (\ref{a-d16_a}),
  \be
\di_\alf {\scJ}{}^\alf(\xi) \equiv \cN_\alf {\scJ}{}^\alf(\xi) \equiv 0\,.
\m{CL_current}
\ee
The current is expressed identically through a divergence of the superpotential, as well,
  \be
{\scJ}{}^\alf(\xi) \equiv \di_\beta {\scJ}{}^{\alf\beta}(\xi) \equiv \cN_\beta {\scJ}{}^{\alf\beta}(\xi)\,.
\m{CL_c_s}
\ee
The related superpotential, see (\ref{a-d22}), is
  \be
{\scJ}{}^{\alf\beta}(\xi) = -{\sM}_\sig{}^{\alf\beta}\xi^\sig =  {{\di \sL} \over {\di  h^a{}_{\beta,\alf}}}h^a{}_\sigma\xi^\sig = \frac{h}{\k}{\sS}_a{}^{\alf\beta}h^a{}_\sigma\xi^\sig\,.
\m{super}
\ee
The direct calculation gives
  \be
{\sJ}{}^{\alf\beta}(\xi) = -\frac{h}{2\k}\l(\cA{}^i{}_{j\mu}- \sA{}^i{}_{j\mu}\r)\l[\delta^\mu_\sigma\l( h_i{}^\alf h^{j\beta} - h_i{}^\beta h^{j\alf}\r) -2
h_i{}^\mu\l(\delta^\alf_\sig h^{j\beta}- \delta^\beta_\sig h^{j\alf} \r) \r]\xi^\sigma\,.
\m{super-bar-bullet}
\ee
  Formally this result coincides with (\ref{super-bar}). However, here $\sA{}^i{}_{j\mu}$ has been introduced into consideration from the start, basing on the inner philosophy of constructing covariant TEGR \cite{Aldrovandi_Pereira_2013}. On the other hand, $\bar A{}^i{}_{j\mu}$ in (\ref{super-bar}) is introduced ``by hand'' at the final stage as an additional assumption.

The relations (\ref{CL_current}) and (\ref{CL_c_s}) are identities only, they do not bring a physical content because up to now the field
equations have not been used. After using the equations (\ref{EM})  the current (\ref{current_cL}) transforms to
 \be
{\scJ}{}^\alf(\xi) = h\l[\stheta_\sig{}^\alf + \Theta_\sigma{}^\alf\r]\xi^\sig
  +\frac{h}{\k}{\sS}_\sig{}^{\alf\rho}\cN_\rho\xi^\sig\,.
 \m{current_c_f}
 \ee
 As a result of using the field equations, the identities (\ref{CL_current}) and (\ref{CL_c_s}) become physically sensible differential conservation laws:
  \bea
\di_\alf {\scJ}{}^\alf(\xi) &=& 0\,,
\m{DiffCL_1}\\
{\scJ}{}^\alf(\xi) &=& \di_\beta {\scJ}{}^{\alf\beta}(\xi)\,.
\m{DiffCL_2}
\eea
All the quantities are spacetime covariant and Lorentz covariant. Consequently the integral conserved quantities can be constructed and they are spacetime covariant and Lorentz covariant as well,
\be
 {\cal P}(\xi) = \int_\Sigma dx^3 \scJ_a{}^0(\xi) = \oint_{\di\Sigma} ds_i \scJ_a{}^{0i}(\xi)\,.
 \m{ICQ_1}
 \ee
Thus, the problem of construction of conserved quantities discussed in Introduction (also in \cite{Krssak}) and remarked around outline of (\ref{ICQ}) is resolved by construction of conservation laws (\ref{DiffCL_1}) and (\ref{DiffCL_2}). {\em They are the main theoretical result of the paper.}

Finalizing the subsection, we again stress the role of the inertial spin connection $\sA{}^i{}_{j\mu}$. Recall that it has no influence both on deriving the field equations (see (\ref{MollerEM}) and (\ref{EM})) and on construction of solutions in TEGR. However, incorporation of $\sA{}^i{}_{j\mu}$ into consideration is necessary to construct (\ref{DiffCL_1}) and (\ref{DiffCL_2}). For a comparison, consider conservation laws (\ref{MCL_1}) and (\ref{MCL_2}) in Moller's consideration where conserved quantities are not Lorentz covariant with absence of $\sA{}^i{}_{j\mu}$. Recall again that equations (\ref{MollerEM}) and (\ref{EM}) are equivalent. Thus, (\ref{MollerEM}), without $\sA{}^i{}_{j\mu}$, can be reformulated into the explicitly Lorentz covariant form (\ref{EM}) by identical transformations with the evident presence of $\sA{}^i{}_{j\mu}$. Keeping in mind this fact, in Appendix \ref{Appendix}, we derive the new conservation laws (\ref{DiffCL_1}) and (\ref{DiffCL_2}) starting from the Einstein equations in the metric form. This demonstrates again, an auxiliary (although necessary) role of $\sA{}^i{}_{j\mu}$ in constructing conserved quantities in TEGR which are simultaneously both coordinate covariant and Lorentz invariant and which allow us to construct well defined charges. At last, on the role of $\sA{}^i{}_{j\mu}$ see discussion in section \ref{Comparison}.

\subsection{Principle of determining inertial spin connection}
\m{Principles}

 Although problems of constructing conserved quantities in the covariant TEGR have been resolved above, the relation between the GR spin connection $\cA{}^i{}_{j\mu}$ and the inertial spin connection $\sA{}^i{}_{j\mu}$ is left undetermined. Now, analyzing the new conserved quantities in the covariant TEGR, we give a rule to define this relation. It is quite natural requirement that in the case of absence of gravity  both the new current and the new superpotential have to vanish. Namely this requirement gives a possibility to define a principle for defining $\sA{}^i{}_{j\mu}$. Viewing the structure of the current, see (\ref{current_c_f}), and the superpotential, see (\ref{super-bar-bullet}), one finds that they essentially depend on the contortion $\sK{}^i{}_{j\mu}$ defined in (\ref{contortion}), indeed, see relations (\ref{Hilbert_div+K}), (\ref{contor}), (\ref{S}) and (\ref{E_M+}).  The gravity is `switched off' when $\sK{}^i{}_{j\mu} = 0$ and $\Theta_{\sig}{}^\alf = 0$, just under these conditions both the new current and the new superpotential become zero.

 To introduce constructive properties for the principle of determining  $\sA{}^i{}_{j\mu}$ we make the following. Derive the curvature tensor
\be
\cR{}^i{}_{j\mu\nu}=\di_\mu \cA{}^i{}_{j\nu} - \di_\nu \cA{}^i{}_{j\mu} + \cA{}^i{}_{k\mu}\cA{}^k{}_{j\nu} - \cA{}^i{}_{k\nu}\cA{}^k{}_{j\mu}\,
\m{curvatureGR}
\ee
that is not zero for a concrete (not flat) physical solution. In the case of `switching off' gravity one has $\cR{}^i{}_{j\mu\nu} \goto 0$.
Now recall that the  curvature tensor for $\sA{}^i{}_{j\mu}$ is zero by definitions in the covariant TEGR, $\sR{}^i{}_{j\mu\nu}=0$.
Then, one rewrites the `switch off' conditions as $\cR{}^i{}_{j\mu\nu} \goto \sR{}^i{}_{j\mu\nu}$. Then it is evidently that under
these conditions we must provide a transformation from $\cA{}^i{}_{j\mu}$ to related (and desirable) $\sA{}^i{}_{j\mu}$, that is
$\cA{}^i{}_{j\mu} \goto \sA{}^i{}_{j\mu}$.

Remark that the notion of  `switching off' gravity already has been used in literature in order to define
$\cA{}^i{}_{j\mu} \goto \sA{}^i{}_{j\mu}$. Thus, for example, in \cite{Obukhov+}, in the case of the Schwarzschild solution it was
suggested to set $M =0$. More generally it is suggested to `switch off' the Newtonian constant $G\goto 0$ in concrete solutions, see, for
example, \cite{Saridakis+}. However, it cannot work everywhere because, for example, it is not necessary that the vacuum solutions with
non-zero curvature contain elements with $G$. Therefore, our suggestion to `switch off' gravity in (\ref{curvatureGR}) by the requirement $\cR{}^i{}_{j\mu\nu}\goto 0$ is universal.

\section{Inertial spin connection and background spacetime}
\setcounter{equation}{0}
\m{Comparison}

To understand more in theoretical principles of constructing new conserved quantities in the covariant TEGR it is useful to compare them with well known methods in the usual metric presentation of GR.

From the beginning, let us repeat briefly the way of constructing conserved quantities based on the Moller Lagrangian. Starting from the Hilbert Lagrangian in the form (\ref{Hilbert}), one selects a divergence in (\ref{Hilbert_div}) and obtains the Moller Lagrangian (\ref{Moller}). The last is left coordinate covariant, but becomes non-invariant with respect to local Lorentz rotations. Applying the Noether theorem to  (\ref{Moller}), one obtains as Lorentz non-covariant all the quantities: energy-momentum  (\ref{MollerTEI}), conserved current  (\ref{current_M+}) and superpotential  (\ref{super_M}).

 A construction of the covariant TEGR Lagrangian (\ref{Hilbert_div+K}), the same (\ref{lag}), follows from the Hilbert Lagrangian in the form (\ref{Hilbert+}) after subtracting a divergence in (\ref{Hilbert_div+}). Unlike the Moller Lagrangian (\ref{Moller}), the TEGR Lagrangian (\ref{lag}) is both Lorentz invariant and coordinate covariant. As a result, application of the Noether theorem leads to constructing Lorentz and coordinate covariant all the quantities: energy-momentum  (\ref{E_M+}), conserved current  (\ref{current_c_f}) and superpotential  (\ref{super}). Keeping in mind the formal properties only, we accent that presentations of the Hilbert Lagrangian in (\ref{Hilbert}) and in (\ref{Hilbert+}) are equivalent ones. Thus, the components of inertial spin connection $\sA^i{}_{j\mu}$ present an external (auxiliary) structure in (\ref{Hilbert+}) because the last after identical transformations transforms to (\ref{Hilbert}).

This picture is quite analogous to the one in metric GR. To construct Lagrangian with only the first derivatives Einstein subtracted a related divergence.
 As a result, he has obtained the well known truncated Einstein Lagrangian \cite{Landau_Lifshitz_1975}. However, it is not covariant with respect
to general coordinate transformations. In spite of that, it is left invariant with respect to constant coordinate displacements. This permits to
apply the Noether theorem that gives a well known energy-momentum complex (Einstein's pseudotensor) and Freud's superpotential, see Chapter 1
in the book \cite{Petrov_KLT_2017}.\footnote{Remark, if one applies the Noether theorem using invariance with respect to constant coordinate displacements only, then even a covariant metric Lagrangian, like the Hilbert Lagrangian $\lag_H \sim \cR$, leads, anyway, to pseudotensors (not tensors) and not covariant superpotentials, like metric Moller's superpotential, for a detail see Chapter 1 in the book \cite{Petrov_KLT_2017}. Recently, in \cite{Capozziello+2018} the theories $f(R)$ and $f(T)$ have been studied and compared. Both of the Lagrangians are Lorentz and coordinate covariant, but the Noether theorem with using invariance with respect to constant coordinate displacements only has been applied. As a result, the authors have obtained conserved pseudotensors only, not tensors.} Both of these quantities are not covariant, the same as the truncated Einstein Lagrangian. Thus, at this level the analogy with the Moller picture is clear. The Moller Lagrangian (\ref{Moller}) is not invariant with respect to local Lorentz rotations and leads to Lorentz non-covariant conserved quantities. On the other hand, the truncated Einstein Lagrangian is not coordinate covariant and leads to coordinate non-covariant conserved quantities.

For a comparison in the metric presentation of GR we choose the method suggested by Katz, Bi\v c\'ak and Lynden-Bell (KBL) in \cite{KBL_1997}
that gives a possibility, making the use of the Noether theorem, to construct conserved
coordinate covariant current and related superpotential that contain as parts covariantized both Einstein's pseudotensor and Freud's superpotential.
Their main idea is to construct a relative bi-metric Lagrangian with incorporating auxiliary background
metric $\bar g_{\mu\nu}$ and related Christoffel symbols $\bar \Gamma^\alf{}_{\beta\gamma}$. The usual form of the
Hilbert Lagrangian in the metric presentation is
\be
\lag_H  = -\frac{1}{2\k} \sqrt{-g}g^{\mu\nu}\l( \di_\rho \cG{}^\rho{}_{\mu\nu} - \di_\nu \cG{}^\rho{}_{\rho\mu} + \cG{}^\rho{}_{\mu\nu}\cG{}^\sig{}_{\sig\rho} - \cG{}^\rho{}_{\sig\mu}\cG{}^\sig{}_{\rho\nu}\r)\,,
\m{Hilbert_metric}
\ee
where $\cG{}^\rho{}_{\mu\nu}$ are the Christoffel symbols compatible with the physical metric $g_{\mu\nu}$, see (\ref{cG}). The background strictures, $\bar g_{\mu\nu}$ and $\bar \Gamma^\alf{}_{\beta\gamma}$, can be incorporated into (\ref{Hilbert_metric}), and the last is rewritten identically in the evidently covariant form
 \be
\lag_H  = -\frac{1}{2\k} \sqrt{-g}g^{\tau\sig}\l(
\bar\nabla_\rho \Delta^\rho{}_{\tau\sig} -  \bar\nabla_\sig\Delta^\rho{}_{\tau\rho} +
 \Delta^\rho{}_{\rho\eta} \Delta^\eta{}_{\tau\sig} -
 \Delta^\rho{}_{\sig\eta} \Delta^\eta{}_{\tau\rho} + \bar R_{\tau\sig}\r)\,.
 \m{e-a5}
 \ee
 Here, the difference
 \be
 \Delta^\rho{}_{\tau\sig} = \cG{}^\rho{}_{\tau\sig} - \bar\Gamma^\rho{}_{\tau\sig}
 \m{Delta}
 \ee
 is a tensor of third rank, and $\bar\nabla_\rho$ is a covariant derivative compatible with $\bar g_{\mu\nu}$. To derive the covariant KBL Lagrangian one has to subtract from the Hilbert Lagrangian (\ref{e-a5}) a related divergence and the background Hilbert Lagrangian. Thus,
 \be
  \lag_{\scriptscriptstyle KBL} = \lag_H - \bar \lag_H - \di_\alf {\hat k}^\alf; \qquad {\hat k}^\alf =   {\sqrt{-g}\over {2\k}}\l(g^{\alpha\nu}\Delta^\mu{}_{\mu\nu} - g^{\mu\nu} \Del^\alpha{}_{\mu\nu} \r)\, .
\m{e-a2}
 \ee
After that one obtains the KBL Lagrangian (\ref{e-a2}) in the final explicit form:
 \be
  \lag_{\scriptscriptstyle KBL} = -{1\over {2\k}} \sqrt{-g}g^{\tau\sig}\l(
 \Delta^\rho{}_{\rho\eta} \Delta^\eta{}_{\tau\sig} - \Delta^\rho{}_{\sig\eta} \Delta^\eta{}_{\tau\rho}\r)-{1\over {2\k}}
 \l(\sqrt{-g}g^{\tau\sig} - \sqrt{-\bar g}\bar g^{\tau\sig}\r)\bar R_{\tau\sig}\,.
\m{e-a7}
 \ee

 Now, we are in a position to compare the method of constructing new conserved quantities in the covariant TEGR suggested here and the KBL method in metric presentation of GR.

 \noindent {\em Analogies:}

 First, both of the methods start from the Hilbert Lagrangian without additional structures a) in the tetrad form (\ref{Hilbert}) and b) in the metric form (\ref{Hilbert_metric}).

 Second, in the covariant TEGR, one defines Lorentz covariant tensor (contortion) in (\ref{contortion}) as a difference of spin connections; in metric GR, one defines a tensor of 3-d rank in (\ref{Delta}) as a difference of Christoffel symbols.

 Third, in the covariant TEGR, one uses the Hilbert Lagrangian in explicitly Lorentz invariant form  (\ref{Hilbert+}) with incorporating external inertial spin connection $\sA{}^i{}_{j\mu}$; in metric GR, one uses the Hilbert Lagrangian in explicitly coordinate covariant form (\ref{e-a5}) with incorporating external structures of a background spacetime, $\bar g_{\mu\nu}$ and  $\bar \Gamma^\alf{}_{\beta\gamma}$.

 Fourth, in the covariant TEGR, one derives the Lorentz invariant Lagrangian of the first derivatives only (\ref{Hilbert_div+K}); in metric GR, one derives the coordinate covariant KBL Lagrangian (\ref{e-a7}) of the first derivatives only.

 Fifth, in the covariant TEGR, the Lagrangian (\ref{Hilbert_div+K}) leads to construction of Lorentz covariant conserved quantities, unlike the Moller Lagrangian; in metric GR the KBL Lagrangian (\ref{e-a7}) leads to construction of coordinate covariant conserved quantities, unlike the truncated Einstein Lagrangian.

 Sixth, in the KBL method, a background spacetime usually is chosen by a concrete problem under consideration; in the covariant TEGR, to choose $\sA{}^i{}_{j\mu}$ we must follow, for example, the principle given in previous subsection that is connected with the solution under consideration as well.

 \noindent {\em Differences:}

 The first remark is a formal one. The KBL Lagrangian (\ref{e-a7}) includes extra term, $\bar\lag_H$, with respect to the covariant TEGR Lagrangian (\ref{Hilbert_div+K}). It is because the KBL method permits to choose an arbitrary curved background including non-vacuum spacetimes with non-zero curvature defined by $\bar \Gamma^\alf{}_{\beta\gamma}$; whereas in the covariant TEGR the curvature defined by $\sA{}^i{}_{j\mu}$ must be zero in all the cases.

 Second, in the KBL method, a background spacetime can be chosen really ``by hand'' because in the mathematical formalism there are no restrictions in such a choice. On the other hand, in the covariant TEGR, to choose $\sA{}^i{}_{j\mu}$ one has to turn to the solution under consideration anyway. This fact, possibly means that the rules for searching for $\sA{}^i{}_{j\mu}$ are incorporated in the inner properties of the formalism of the covariant TEGR.

 \section{Mass of the Schwarzschild black hole}
\setcounter{equation}{0}
\m{mass}

In this section, we use the results of subsection \ref{TEGR_CL} to obtain mass of the Schwarzschild black hole as a conserved charge (\ref{ICQ_1}).

\subsection{The conserved charge}
\m{charge}

Consider metric in the spherical Schwarzschild coordinates:
\begin{equation} \label{BHmet}
ds^{2} =-\left(1-\frac{2M}{r} \right)dt^{2} +\left(1-\frac{2M}{r} \right)^{-1} dr^{2} +r^{2} \left(d\theta ^{2} +\sin ^{2} \theta d\phi ^{2} \right)
\end{equation}
Recall that the dynamical variables in TEGR introduced in (\ref{g_hh_GR}) as the tetrad components $h^a {}_{\mu}$ admit arbitrary local Lorentz transformation. So, we can choose every tetrad satisfying metric obtained by solving field equations of GR. Usually it is more convenient to work with a diagonal tetrad because one can apply simple transformations between coordinate and local indices. So, from the start we construct the diagonal tetrad from the metric (\ref{BHmet}), keeping in mind  (\ref{g_hh_GR}),
\begin{equation} \label{BHtet}
h^{a} {}_{\mu } =\left[\begin{array}{cccc} {\left(1-\frac{2M}{r} \right)^{{1\mathord{\left/ {\vphantom {1 2}} \right. \kern-\nulldelimiterspace} 2} } } & {0} & {0} & {0} \\ {0} & {\left(1-\frac{2M}{r} \right)^{{-1\mathord{\left/ {\vphantom {-1 2}} \right. \kern-\nulldelimiterspace} 2} } } & {0} & {0} \\ {0} & {0} & {r} & {0} \\ {0} & {0} & {0} & {r\sin \theta } \end{array}\right]\, .
\end{equation}

First we calculate GR spin connection (\ref{circ_A}) for the metric (\ref{BHmet}) and the tetrad (\ref{BHtet}). The non-zero components are:
\begin{equation} \label{BHGRspin}
\begin{array}{l} {\stackrel{\circ}{A} {}^{0} {}_{10} =\stackrel{\circ}{A} {}^{1} {}_{00} =\frac{M}{r^{2} } ;{\mathop{}\nolimits^{}} \stackrel{\circ}{A} {}^{1} {}_{22} =-\stackrel{\circ}{A} {}^{2} {}_{12} =-\left(1-\frac{2M}{r} \right)^{{1\mathord{\left/ {\vphantom {1 2}} \right. \kern-\nulldelimiterspace} 2} } ;} \\ {\stackrel{\circ}{A} {}^{1} {}_{33} =-\stackrel{\circ}{A} {}^{3} {}_{13} =-\sin \theta \left(1-\frac{2M}{r} \right)^{{1\mathord{\left/ {\vphantom {1 2}} \right. \kern-\nulldelimiterspace} 2} } ;{\mathop{}\nolimits^{}} \stackrel{\circ}{A} {}^{2} {}_{33} =-\stackrel{\circ}{A} {}^{3} {}_{23} =-\cos \theta .} \end{array}
\end{equation}
To define inertial spin connection $\stackrel{\bullet}{A} {}^{a} {}_{c\mu }$ we need to calculate $\stackrel{\circ}{A} {}^{a} {}_{c\mu }$ at the ``background'' where the gravity is off, i.e.,  when the Riemannian tensor vanishes: $\stackrel{\circ}{R} {}^a {}_{b \gamma \delta}=0$. It is obvious that $M = 0$ solves this equation. Then, in absence of gravity the components of the GR spin connection $\stackrel{\circ}{A}{}^{a} {}_{c\mu }$ in (\ref{BHGRspin}) goes to the components of the inertial spin connection $\stackrel{\bullet}{A} {}^{a} {}_{c\mu }$ related to the black hole:
\begin{equation} \label{BHspin}
\stackrel{\bullet}{A} {}^{1} {}_{22} =-\stackrel{\bullet}{A} {}^{2} {}_{12} =-1;{\mathop{}\limits^{}} \stackrel{\bullet}{A} {}^{1} {}_{33} =-\stackrel{\bullet}{A} {}^{3} {}_{13} =-\sin \theta ;{\mathop{}\limits^{}} \stackrel{\bullet}{A} {}^{2} {}_{33} =-\stackrel{\bullet}{A} {}^{3} {}_{23} =-\cos \theta .
\end{equation}
Now, using (\ref{contortion}), we get contortion $\stackrel{\bullet}{K} {}^{a} {}_{c\mu }$, the non-zero components of which are:
\begin{equation} \label{BHcont}
\begin{array}{l} {\stackrel{\bullet}{K} {}^{0} {}_{10} =\stackrel{\bullet}{K} {}^{1} {}_{00} =-\frac{M}{r^{2} } ;{\mathop{}\nolimits^{}} \stackrel{\bullet}{K} {}^{1} {}_{22} =-\stackrel{\bullet}{K} {}^{2} {}_{12} =\left(1-\frac{2M}{r} \right)^{{1\mathord{\left/ {\vphantom {1 2}} \right. \kern-\nulldelimiterspace} 2} } -1;} \\ {\stackrel{\bullet}{K} {}^{1} {}_{33} =-\stackrel{\bullet}{K} {}^{3} {}_{13} =\sin \theta \left[\left(1-\frac{2M}{r} \right)^{{1\mathord{\left/ {\vphantom {1 2}} \right. \kern-\nulldelimiterspace} 2} } -1\right]}\,. \end{array}
\end{equation}
Then the torsion tensor $\stackrel{\bullet}{T} {}^{\alpha} {}_{\mu \nu}$, which is antisymmetric in lower indices, is defined by (\ref{tor+}). The non-zero components of it are:
\begin{equation} \label{BHtor}
\stackrel{\bullet}{T} {}^{0} {}_{10} =\frac{M}{r^{2} } \left(1-\frac{2M}{r} \right)^{-1} ;{\mathop{}\nolimits^{}} \stackrel{\bullet}{T} {}^{2} {}_{12} =\stackrel{\bullet}{T} {}^{3} {}_{13} =\frac{1}{r} \left[1-\left(1-\frac{2M}{r} \right)^{{1\mathord{\left/ {\vphantom {1 2}} \right. \kern-\nulldelimiterspace} 2} } \right]
\end{equation}
and the same with opposite sign for swapped lower indices. Non-zero components of the tensor $\sS{}_{\mu}{}^{\rho\sigma} = \sS{}_{a}{}^{\rho\sigma}h^a{}_\mu $ defined by  (\ref{S}), which are antisymmetric in upper indices, are:
\begin{equation} \label{BHsup}
\stackrel{\bullet}{S}_{0} {}^{01} =\frac{2}{r} \left[1-\left(1-\frac{2M}{r} \right)^{{1\mathord{\left/ {\vphantom {1 2}} \right. \kern-\nulldelimiterspace} 2} } \right]-\frac{4M}{r^{2} } ;{\mathop{}\limits^{}} \stackrel{\bullet}{S}_{2} {}^{12} =\stackrel{\bullet}{S}_{3} {}^{13} =-\frac{1}{r} \left[1-\left(1-\frac{2M}{r} \right)^{{1\mathord{\left/ {\vphantom {1 2}} \right. \kern-\nulldelimiterspace} 2} } \right]+\frac{M}{r^{2} }\,,
\end{equation}
and the same with opposite sign for swapped upper indices.

Now we are in a position to calculate the total mass/energy of the Schwarzschild black hole. We use (\ref{super}) and (\ref{ICQ_1}), and choose a displacement vector as the time-like Killing vector $\xi ^{\alpha } =\left(-1,0,0,0\right)$. As a result, we obtain
\begin{equation} \label{GrindEQ__35}
E=\frac{1}{\kappa } {\mathop{\lim }\limits_{r\to \infty }} \int _{\partial \Sigma }dx^{2}  h\stackrel{\bullet}{S}_{0} {}^{01} \xi ^{0} =M
\end{equation}
where it was used $\kappa =8\pi $ , $h=r^{2} \sin \theta $.

Recall that the charge defined in (\ref{ICQ_1}) does not depend on local Lorentz rotations by the construction. So, after arbitrary Lorentz rotations the integration in (\ref{GrindEQ__35}) gives the same result, $E=M$.  By a related
Lorentz rotation one can suppress $\stackrel{\bullet}{A} {}^{a} {}_{c\mu } $ totally: $\stackrel{\bullet}{A} {}^{a} {}_{c\mu } =0$ that is to use Weitzenb\"ock gauge. Let us transform to that. Applying (\ref{lroth}) and (\ref{lrotA}) to (\ref{BHtet}) and (\ref{BHspin}) with the Lorentz matrix:
\begin{equation} \label{GrindEQ__36_}
\Lambda ^{a} {}_{b} =\left[\begin{array}{cccc} {1} & {0} & {0} & {0} \\ {0} & {\sin \theta \cos \phi } & {\cos \theta \cos \phi } & {-\sin \phi } \\ {0} & {\sin \theta \sin \phi } & {\cos \theta \sin \phi } & {\cos \phi } \\ {0} & {\cos \theta } & {-\sin \theta } & {0} \end{array}\right]\,,
\end{equation}
the tetrad components become
\begin{equation} \label{GrindEQ__37_}
h^{a} {}_{\mu } =\left[\begin{array}{cccc} {\left(1-\frac{2M}{r} \right)^{{1\mathord{\left/ {\vphantom {1 2}} \right. \kern-\nulldelimiterspace} 2} } } & {0} & {0} & {0} \\ {0} & {\sin \theta \cos \phi \left(1-\frac{2M}{r} \right)^{{-1\mathord{\left/ {\vphantom {-1 2}} \right. \kern-\nulldelimiterspace} 2} } } & {r\cos \theta \cos \phi } & {-r\sin \theta \sin \phi } \\ {0} & {\sin \theta \sin \phi \left(1-\frac{2M}{r} \right)^{{-1\mathord{\left/ {\vphantom {-1 2}} \right. \kern-\nulldelimiterspace} 2} } } & {r\cos \theta \sin \phi } & {r\sin \theta \cos \phi } \\ {0} & {\cos \theta \left(1-\frac{2M}{r} \right)^{{-1\mathord{\left/ {\vphantom {-1 2}} \right. \kern-\nulldelimiterspace} 2} } } & {-r\sin \theta } & {0} \end{array}\right]\,,
\end{equation}
whereas the inertial spin connection vanishes, $\sA{}^a {}_{b \mu}=0$. Using this gauge in (\ref{GrindEQ__35}), we obtain again $E=M$.

\subsection{Discussion}
\m{discourse}

It is useful to compare our results with those in a more earlier paper \cite{M_2M}, where the Moller presentation of GR, with the Lagrangian (\ref{Moller}), equations (\ref{MollerEM}) simplified to the vacuum state, energy-momentum (\ref{MollerTEI}) and superpotential (\ref{a-d9_M}), is studied. There is no a presence of $\sA{}^i{}_{j\mu}$ totally. The authors have found two different spherically symmetrical static vacuum tetrad solutions.  They correspond to two known metrics of the Schwarzschild solution that are connected by related coordinate transformations. The calculation of the total charge with making the use of the superpotential for the first tetrad gives unacceptable $2M$, whereas for the second tetrad one obtains acceptable $M$. The authors cannot explain this result. But in later studies (see, for example, \cite{Obukhov+}, and our results here) one explains the problem of \cite{M_2M} easily. It turns out that both of the tetrad solutions in \cite{M_2M} are connected not only by coordinate transformations, but by a local Lorentz rotation as well. Therefore, because the Moller presentation is not Lorentz covariant one obtains different charges. The case of acceptable result $M$ corresponds to a so-called {\em proper tetrad} in {\em Weitzenb\"ock gauge} both in our presentation here and in \cite{Obukhov+}.

Now, it is instructively to discuss results of the paper \cite{Obukhov+} in a more detail. The authors consider inertial spin connection as a regularizer in constructing conserved quantities.  From the start they notice that if one sets $\sA{}^i{}_{j\mu}=0$ then the relations (\ref{EM_Lc}) and (\ref{EM+}) coincide and constructing conserved charges becomes possible in both the cases. On the example of constructing the total mass for the Schwarzschild and Kerr black holes they demonstrate principles of the regularization. We turn to the Schwarzschild case only. First, one derives a diagonal tetrad related to the Schwarzschild metric in the Schwarzschild coordinates. With related superpotential one tries to calculate the mass by the formula (\ref{ICQ}) with $\sA{}^i{}_{j\mu}=0$.  As a result, one obtains a divergent integration. To compensate the divergence one introduces a related inertial spin connection $\sA{}^i{}_{j\mu}\neq 0$ defined as $\sA{}^i{}_{j\mu} = \l.\cA{}^i{}_{j\mu}\r|_{M=0} $ and obtains acceptable mass $M$. {\em It is a suggested regularization.} After that
 a local Lorentz rotation that suppresses the introduced inertial spin connection is defined. Recall that it is called the {\em Weitzenb\"ock gauge} and the related tetrad in this case is called the {\em proper tetrad}. Probably, this notion has been introduced in the first time. Under local Lorentz rotations the GR spin connection as well as the tetrad are changed, but the result $M$ is not changed. Namely in this case of the proper tetrad the Moller presentation can be used and a consistent construction of charges becomes permissible. The difference from our approach is, first, that we take into account $\sA{}^i{}_{j\mu}$ from the start as a necessary structure of the formalism in whole, whereas in \cite{Obukhov+} it is introduced as a compensating structure in constructing conserved quantities. Second, taking into account a displacement vector in all the expressions explicitly, we can construct charges in an arbitrary gauge, whereas in \cite{Obukhov+} it is necessary the Weitzenb\"ock gauge.

 At last, in \cite{f(T)5}, under the requirement of a consistence of field equations in $f(T)$ theories the authors are searching proper tetrads, or appropriate inertial spin connections. Essential condition is that symmetries of assumed solutions are taken into account. In the case of spherical symmetries their proper tetrad adopted to the Schwarzschild solution is simplified exactly to our result (\ref{GrindEQ__37_}). It surprisingly, but it turns out that quite different principles lead to the same results.

\section{A freely falling observer}
\setcounter{equation}{0}
\m{observer}

Here, we calculate densities of conserved quantities for a freely falling observer in the FLRW universe and in the (A)dS space. To follow this goal we need to find the components of the Noether current ${\scJ}{}^{\alf } (\xi )$ defined in (\ref{current_c_f}) with a proper vector, $\xi^\alf$, of the observer. To understand better what quantities will be calculated let us turn to special relativity. Consider the metric element of the Minkowski space in the form:
\begin{equation}\label{metMin}
    ds^2=-dt^2+dr^2+r^2\l(d\theta^2+ \sin^2\theta  d\phi^2\r)\,.
\end{equation}
Let the observer be at rest in the frame of (\ref{metMin}) with the proper vector $\xi^\alf = (-1,0,0,0)$; and let the symmetric energy-momentum tensor, $T_{\mu\nu}$, be introduced for matter propagating in the Minkowski space. After that define a conserved current ${\cal I}^\alf = T^\alf{}_\beta \xi^\beta$. Then ${\cal I}^0$ can be interpreted as the energy density measured by the observer. In this simplest case ${\cal I}^i$ can be interpreted as momentum density. Under arbitrary coordinate transformations components ${\cal I}^\alf $ are transformed   as vector ones with a correspondence to the tensorial law. In the case ${\cal I}^\alf =0 $, one states the absence of densities of conserved quantities, like energy and momentum, measured by the observer. Keeping in mind the above simple notions, we will interpret the components ${\scJ}{}^{\alf } (\xi )$.
 Note also that, following the conservation law (\ref{DiffCL_2}):
    \begin{equation}
{\scJ}{}^{\alpha } (\xi ) = \partial _{\beta } {\scJ}{}^{\alpha \beta } (\xi ),
\m{CL_FLRW}
\end{equation}
one can check the current components expressed through a divergence of the superpotential, ${\scJ}{}^{\alpha \beta } (\xi )$, defined in (\ref{super}).

\subsection{The FLRW universe}
\m{FLRW}

We consider the FLRW metric in the form:
\begin{equation}\label{metfrid}
    ds^2=-dt^2+a^2(t)\l(\frac{1}{1-kr^2}dr^2+r^2 \l(d\theta^2+ \sin ^2\theta  d\phi^2 \r)\r)
\end{equation}
where $k=+1$ for a positively curved space, $k=0$ for a flat space and $k=-1$ for a negatively curved space. Then, following (\ref{g_hh_GR}), we take a more convenient for calculations diagonal tetrad,
\begin{equation}
    h^a{}_\mu={\diag} \l(1,{a}/{\chi},a r,a r \sin \theta \r)\,.
    \m{tetradFLRW}
\end{equation}
Here and below we denote $\chi = \sqrt{1-kr^2}$; for the sake of definiteness we chose the positive sign of $\chi$ only; because we study local characteristics only we require that $\chi$ requires real values only.
Following (\ref{circ_A}), we calculate components of the GR spin connection  for (\ref{metfrid}) and (\ref{tetradFLRW}):
\begin{equation}\label{friedlevispin}
\begin{array}{cccc}

{\cA}{}^0{}_{11} ={\cA}{}^1{}_{01} ={\dot{a}}/{\chi}
~,~
{\cA}{}^0{}_{22} ={\cA}{}^2{}_{02} =\dot{a} r
~,~
{\cA}{}^0{}_{33} ={\cA}{}^3{}_{03} =\dot{a} r \sin \theta \,,

\\
{\cA}{}^1{}_{22} =-{\cA}{}^2{}_{12} =-{\chi} ~,~
{\cA}{}^1{}_{33} =-{\cA}{}^3{}_{13} =-{\chi}\sin \theta  \,,
\\
{\cA}{}^2{}_{33} =-{\cA}{}^3{}_{23} =-\cos\theta\,.

\end{array}
\end{equation}

To calculate the inertial spin connection, again we need to solve  $\stackrel{\circ}{R} {}^a {}_{b \gamma \delta} =0$. For the system under consideration it is more appropriate the equivalent equation ${\cR}{}^a {}_{b c d} = {\cR} {}^a {}_{b \gamma \delta}h_c{}^\gamma h_d{}^\delta =0$. Thus, all the non-zero
tetrad components $\stackrel{\circ}{R} {}^a {}_{b c d} $ are only
\begin{equation}\label{fridriem}
R^0 {}_{i 0 i} =-R^0 {}_{i i0} =R^i {}_{0 0 i} =-R^i {}_{0 i0} = \frac{\ddot{a}}{a}
,~~~~
R^i {}_{j i j} =- R^i {}_{j ji} = \frac{k+\dot{a}^2}{a^2}
\end{equation}
where $i,j = 1,2,3$. Equating these components to zero we got two equations: $\dot{a}^2+k =0$ and $\ddot{a}=0$.
All solutions to the first equation satisfy the second equation, so only
\begin{equation}\label{fridvaceq}
\dot{a}^2+k =0
\end{equation}
is important. It is interesting to note that solving ``vacuum'' Friedman equation
  \begin{equation}
      H^2=\rho_{curv}\qquad {\rm or}\qquad \left( \frac{\dot{a}}{a}\right)^2=-\frac{k}{a^2}\,,
      \m{H_vac}
  \end{equation}
one obtains (\ref{fridvaceq}) as well.

Taking the solution to (\ref{fridvaceq}) in the united form $a(t)= \sqrt{-k} t$ for all the three signs of curvature and substituting it into the GR
spin connection (\ref{friedlevispin}), we get the non-zero components of inertial spin connection:
\begin{equation}
\begin{array}{cccc}

\stackrel{\bullet}{A}{}^0{}_{11} =\stackrel{\bullet}{A}{}^1{}_{01} ={\sqrt{-k}}/{\chi}
~,~
\stackrel{\bullet}{A}{}^0{}_{22} =\stackrel{\bullet}{A}{}^2{}_{02} =\sqrt{-k} r
~,~
\stackrel{\bullet}{A}{}^0{}_{33} =\stackrel{\bullet}{A}{}^3{}_{03} =\sqrt{-k} r \sin\theta\,,

\\
\stackrel{\bullet}{A}{}^1{}_{22} =-\stackrel{\bullet}{A}{}^2{}_{12} =-\chi ~,~
\stackrel{\bullet}{A}{}^1{}_{33} =-\stackrel{\bullet}{A}{}^3{}_{13} =-\chi\sin\theta \,,
\\
\stackrel{\bullet}{A}{}^2{}_{33} =-\stackrel{\bullet}{A}{}^3{}_{23} =-\cos\theta \,.

\end{array}
\m{inertialSCFLRW}
\end{equation}
For each of the signs of curvature one can set $\sqrt{-k}= \pm i; 0; \pm 1 $. Thus, the components (\ref{inertialSCFLRW}) become complex values\footnote{At the end of the section, we discuss possibilities to use complex components of the spin connection.} for $k=+1$; are related to Minkowski space $a(t)=\const$ for $k=0$; and are related to the Milne solution, $a(t)=\pm t$ for $k=-1$.

For the system under consideration it is more economical to present the components of the contortion tensor in the tetrad components $    \stackrel{\bullet}{K}{}^a{}_{b c} $:
\be
\stackrel{\bullet}{K}{}^0 {}_{11} =\stackrel{\bullet}{K}{}^1{}_{01} = \stackrel{\bullet}{K}{}^0 {}_{22} =\stackrel{\bullet}{K}{}^2{}_{02} = \stackrel{\bullet}{K}{}^0 {}_{33} =\stackrel{\bullet}{K}{}^3{}_{03}=-H+\sqrt{-k}/a\,.
\m{K_FLRW}
\ee
The spacetime components of the tensors $    \stackrel{\bullet}{T}{}^\alpha{}_{\beta \mu} $ and $    \stackrel{\bullet}{S}{}_\alpha{}^{\beta \mu} $ are respectively,
\be
\stackrel{\bullet}{T}{}^1 {}_{10} =-\stackrel{\bullet}{T}{}^1{}_{01} =\stackrel{\bullet}{T}{}^2 {}_{20} =-\stackrel{\bullet}{T}{}^2{}_{02} =\stackrel{\bullet}{T}{}^3 {}_{30} =-\stackrel{\bullet}{T}{}^3{}_{03} =-H+\sqrt{-k}/a\,,
\m{T_FLRW}
\ee
\be
\stackrel{\bullet}{S}{}_1 {}^{10} =-\stackrel{\bullet}{S}{}_1{}^{01} =\stackrel{\bullet}{S}{}_2 {}^{20} =-\stackrel{\bullet}{S}{}_2{}^{02} =\stackrel{\bullet}{S}{}_3 {}^{30} =-\stackrel{\bullet}{S}{}_3{}^{03} = -2(H-\sqrt{-k}/a)\,.
\m{S_FLRW}
\ee
At last, we consider an observer at the rest with respect to co-moving coordinates of the metric (\ref{metfrid}). Then components of his proper vector are
\be
\xi^\sigma=(-1,0,0,0)\,.
\m{proper_FLRW}
\ee

Now, we are in a position to calculate components of the current (\ref{current_c_f}):
\be
\scJ {}^{\alf } (\xi ) =h\l(\stheta_\sig{}^\alf +\Theta_\sig{}^\alf\r)\xi^\sig + {h}{\kappa }^{-1} \sS {}_{a} {}^{\alf \rho } h^{a} {}_{\sigma } {\mathop{\nabla _{\rho } }\limits^{\circ }} \xi ^{\sigma }.
\m{current_0}
\ee
By (\ref{proper_FLRW}), we need the components $\stheta_0{}^\alf$ and $\Theta_0{}^\alf$ only. Among them only the 00-components are non-zero's. Thus, by (\ref{E_M+}) one has $\stheta_0 {}^0 =-3\kappa^{-1} (H-\sqrt{-k}/a)^2$. Keeping in mind 00-component of the Friedmann equations \cite{Landau_Lifshitz_1975}, one has $\Theta_0{}^0 =-3 \kappa^{-1} (H^2+{k}/{a^2})$. Besides, only 0-component of the last term in (\ref{current_0}) is non-zero's. Taking into account (\ref{metfrid}), (\ref{S_FLRW}) and (\ref{proper_FLRW}), one has
 \be
\sS {}_{a} {}^{0 \rho } h^{a} {}_{\sigma } {\mathop{\nabla _{\rho } }\limits^{\circ }} \xi ^{\sigma } = -6 H(H-\sqrt{-k}/a)\,,
 \m{last_FLRW}
 \ee
and finally we got
 \be
 \scJ {}^{\alf } (\xi ) =0\,.
 \m{J=0}
 \ee
 This result is suported by a direct calculation of the right hand side of (\ref{CL_FLRW}). Taking into account (\ref{tetradFLRW}), (\ref{S_FLRW}) and (\ref{proper_FLRW}), one finds $\scJ {}^{\alf } (\xi ) = {\kappa }^{-1}  \partial _{\beta } ( h\sS_{a} {}^{\alf \beta } h^{a} {}_{\sigma } \xi ^{\sigma }) =0$. One can easily recognize this result observing (\ref{S_FLRW}) where components with low 0-component are absent. The result (\ref{J=0}) means that the freely falling observer with the proper vector (\ref{proper_FLRW}) does not measure energy and momentum densities.

To go to the Wietzenb\"ock gauge one has to apply the
Lorentz rotation:
\be
\Lambda^a {}_b =
\left(
\begin{array}{cccc}
 \chi & -\sqrt{-k} r & 0 & 0 \\
 -\sqrt{-k} r \sin\theta \cos\varphi &  \chi \sin\theta \cos\varphi & \cos\theta \cos\varphi & -\sin\varphi \\
 -\sqrt{-k} r \sin\theta \sin\varphi &  \chi \sin\theta \sin\varphi & \cos\theta \sin\varphi & \cos\varphi \\
 -\sqrt{-k} r \cos\theta &  \chi \cos\theta & -\sin\theta & 0 \\
\end{array}
\right)
\m{Rotation_FLRW}
\ee
that, indeed, suppresses the spin connection (\ref{inertialSCFLRW}) totally. The tetrad (\ref{tetradFLRW}) takes the form of the proper tetrad:
\begin{equation}
h^a{}_\mu =
\left(
\begin{array}{cccc}
 \chi & -{\sqrt{-k} r a(t)}/{\chi} & 0 & 0 \\
 -\sqrt{-k} r \sin\theta \cos\varphi & a(t) \sin\theta \cos\varphi & r a(t) \cos\theta \cos\varphi & -r a(t) \sin\theta \sin\varphi \\
 -\sqrt{-k} r \sin\theta \sin\varphi & a(t) \sin\theta \sin\varphi & r a(t) \cos\theta \sin\varphi & r a(t) \sin\theta \cos\varphi \\
 -\sqrt{-k} r \cos\theta & a(t) \cos\theta & -r a(t) \sin\theta & 0 \\
\end{array}
\right)\,.
\m{proptet_FLRW}
\end{equation}
Notice that the Lorentz rotation (\ref{Rotation_FLRW}) can be represented by an action of two matrices  $\Lambda=\Lambda_{rot} \Lambda_{boost}$ where $\Lambda_{rot}$ presents 3 dimensional space rotation already is defined in (\ref{GrindEQ__36_}), and $\Lambda_{boost}$ presents the boost:
\begin{equation}
(\Lambda_{boost} ){}^a {}_b =
\left(
\begin{array}{cccc}
 \chi & -\sqrt{-k} r & 0 & 0 \\
 -\sqrt{-k} r & \chi & 0 & 0 \\
 0 & 0 & 1 & 0 \\
 0 & 0 & 0 & 1 \\
\end{array}
\right)\,.
\m{prop_boost}
\end{equation}

Without doubts, we could provide all the above calculations in the Wietzenb\"ock gauge and finally obtain the same result (\ref{J=0}). It is because all the tensors written in all spacetime indices are invariant under Lorentz rotations; all the tetrad indices are contracted in the expression for the current. So, we obtain again that freely falling observer measures zero's densities of energetic characteristics.

\subsection{The (anti-)de Sitter space}
\m{AdS}

Now, let us consider the (A)dS solution with the metric:
\begin{equation} \label{DSmet}
ds^{2} =-\left(1-\frac{1}{3} \Lambda r^2 \right)dt^{2} +\left(1-\frac{1}{3} \Lambda r^2 \right)^{-1} dr^{2} +r^{2} \left(d\theta ^{2} +\sin ^{2} \theta d\phi ^{2} \right)\,.
\end{equation}
The related diagonal tetrad that is a more convenient one is
\begin{equation} \label{DStet}
h^{a} {}_{\mu } =\left[\begin{array}{cccc} {\left(1-\frac{1}{3} \Lambda r^2 \right)^{{1\mathord{\left/ {\vphantom {1 2}} \right. \kern-\nulldelimiterspace} 2} } } & {0} & {0} & {0} \\ {0} & {\left(1-\frac{1}{3} \Lambda r^2 \right)^{{-1\mathord{\left/ {\vphantom {-1 2}} \right. \kern-\nulldelimiterspace} 2} } } & {0} & {0} \\ {0} & {0} & {r} & {0} \\ {0} & {0} & {0} & {r\sin \theta } \end{array}\right] .
\end{equation}
One recognizes easily that calculations are to be analogous to the ones in the case of the black hole, but with the replacement of $\frac{2 M}{r}$ to $\frac{1}{3} \Lambda r^2$, see (\ref{BHmet}) and (\ref{BHtet}). Following (\ref{circ_A}), we calculate for (\ref{DSmet}) and (\ref{DStet}) components of the GR spin connection, non-zero's of them are
\begin{equation} \label{DSGRspin}
\begin{array}{l} {\stackrel{\circ}{A} {}^{0} {}_{10} =\stackrel{\circ}{A} {}^{1} {}_{00} = - \frac{1}{3}\Lambda r ;{\mathop{}\nolimits^{}} \stackrel{\circ}{A} {}^{1} {}_{22} =-\stackrel{\circ}{A} {}^{2} {}_{12} =-\left(1-\frac{1}{3} \Lambda r^2 \right)^{{1\mathord{\left/ {\vphantom {1 2}} \right. \kern-\nulldelimiterspace} 2} } ;} \\ {\stackrel{\circ}{A} {}^{1} {}_{33} =-\stackrel{\circ}{A} {}^{3} {}_{13} =-\sin \theta \left(1-\frac{1}{3} \Lambda r^2  \right)^{{1\mathord{\left/ {\vphantom {1 2}} \right. \kern-\nulldelimiterspace} 2} } ;{\mathop{}\nolimits^{}} \stackrel{\circ}{A} {}^{2} {}_{33} =-\stackrel{\circ}{A} {}^{3} {}_{23} =-\cos \theta .} \end{array}
\end{equation}
The Riemann tensor for the (A)dS space is derived as usual $\stackrel{\circ}{R} {}^a {}_{b \gamma \delta} = \Lambda\l(h^a{}_\gamma h_{b\delta} - h^a{}_\delta h_{b\gamma} \r)$. We repeat: to calculate the inertial spin connection we need to solve  $\stackrel{\circ}{R} {}^a {}_{b \gamma \delta} =0$ that means that we need to put $\Lambda=0$ in (\ref{DSGRspin}). Then we  obtain
\begin{equation} \label{DSspin}
\stackrel{\bullet}{A} {}^{1} {}_{22} =-\stackrel{\bullet}{A} {}^{2} {}_{12} =-1;{\mathop{}\limits^{}} \stackrel{\bullet}{A} {}^{1} {}_{33} =-\stackrel{\bullet}{A} {}^{3} {}_{13} =-\sin \theta ;{\mathop{}\limits^{}} \stackrel{\bullet}{A} {}^{2} {}_{33} =-\stackrel{\bullet}{A} {}^{3} {}_{23} =-\cos \theta
\end{equation}
that are the same inertial spin connection components as for the black hole (\ref{BHspin}), and that is not surprising.

The components (\ref{DSGRspin}) and (\ref{DSspin}) are written in the mixed indices $\cA{}^a{}_{b\mu}$ and $\sA{}^a{}_{b\mu}$ as usual. On the other hand, we derive out the components of the tensors $\sK{}^{a} {}_{bc}$, $\sT{}^{a} {}_{bc}$ and $\sS{}_{a} {}^{bc}$ in the tetrad indices only that is more economical in the case under consideration. Thus,
\begin{equation}
\stackrel{\bullet}{K} {}^{0} {}_{10} =\stackrel{\bullet}{K} {}^{1} {}_{00} =\frac{{\textstyle \frac{1}{3}}\Lambda r}{\l({1- {\textstyle \frac{1}{3}}\Lambda r^2}\r)^{1/2}}; {\mathop{}\limits^{}} \stackrel{\bullet}{K} {}^{1} {}_{22} =-\stackrel{\bullet}{K} {}^{2} {}_{12}= \stackrel{\bullet}{K} {}^{1} {}_{33} =-\stackrel{\bullet}{K} {}^{3} {}_{13} = \frac{\l({1- {\textstyle \frac{1}{3}}\Lambda r^2}\r)^{1/2}-1}{ r} \,  ,
\m{K_AdS}
\end{equation}
\be
\stackrel{\bullet}{T} {}^{0} {}_{01} = -\stackrel{\bullet}{T} {}^{0} {}_{10} = \frac{{\textstyle \frac{1}{3}}\Lambda r}{\l({1- {\textstyle \frac{1}{3}}\Lambda r^2}\r)^{1/2}};{\mathop{}\limits^{}} \stackrel{\bullet}{T} {}^{2} {}_{12}= \stackrel{\bullet}{T} {}^{3} {}_{13} =-\stackrel{\bullet}{T} {}^{3} {}_{31} =-\stackrel{\bullet}{T} {}^{2} {}_{21} = \frac{\l({1- {\textstyle \frac{1}{3}}\Lambda r^2}\r)^{1/2}-1}{ r}\,,
\m{T_AdS}
\ee
\bea
\stackrel{\bullet}{S} {}_{0} {}^{01} &=&-\stackrel{\bullet}{S} {}_{0} {}^{10} = - {\frac{2}{r}  + \frac{2}{r}\l({1- {\textstyle \frac{1}{3}}\Lambda r^2}\r)^{1/2}  }; \nonumber\\{\mathop{}\limits^{}} \stackrel{\bullet}{S} {}_{2} {}^{12} &=&-\stackrel{\bullet}{S} {}_{2} {}^{21}= \stackrel{\bullet}{S} {}_{3} {}^{13} =-\stackrel{\bullet}{S} {}_{3} {}^{31} = -\frac{1}{r} - \frac{1}{r}{\l({1- {\textstyle \frac{1}{3}}\Lambda r^2}\r)^{-1/2}}  +\frac{2}{r} {\l({1- {\textstyle \frac{1}{3}}\Lambda r^2}\r)^{1/2}}  \,.
\m{S_AdS}
\eea

Now, we derive the components of a proper vector of freely falling observers for (\ref{DSmet}) in the most general form:
\begin{equation}
\xi^\alf = \left [- {C}\l(1- {\textstyle \frac{1}{3}}\Lambda r^2 \r)^{-1}, - \l(C^2 - 1+ {\textstyle \frac{1}{3}}\Lambda r^2 \r)^{1/2},0,0\right ]\,
\m{proper_AdS}
\end{equation}
where $C$ is a constant of the integration of the system of equations for geodesics that can be an arbitrary real quantity. Calculating components of the current $\scJ {}^\alf (\xi )$ and repeating all the steps in the FLRW case we obtain that the 0-component of the current (\ref{current_0}) is
 \be
 \scJ {}^0 (\xi ) =\frac{2}{\k }\frac{h}{r^2} C\left( {\left(1- {\textstyle \frac{1}{3}}\Lambda r^2\right)^{-3/2}}-1\right)\,,
 \m{J_0_AdS}
 \ee
 whereas other components are zero's, $\scJ {}^i (\xi ) = 0$. This result is supported by a direct calculation of the right hand side of (\ref{CL_FLRW}). Among all the free falling observers defined by (\ref{proper_AdS}) we can distinguish a one being at rest with respect to the Hubble flow related to the maximally symmetric (A)dS space. Recall that the Hubble flow has zero velocity at the coordinate origin, $r=0$. Then, placing the observer at the point $r=0$ and choosing $C=1$ in (\ref{proper_AdS}) one has for its proper vector $\xi^\alf = (-1,0,0,0)$, compare with (\ref{proper_FLRW}). First, this means that such an observer is ``frozen'' in the Hubble flow because it is at rest in the coordinate origin, second, the component of the current (\ref{J_0_AdS}) becomes zero, that is now totally $\scJ {}^\alf (\xi ) = 0$. All of these mean that interpretations of densities of conserved quantities for such an observer have to be the same as those for the observer at rest with respect to the Hubble flow in the FLRW case, see discussion in the next subsection.

 Finalizing subsection, let us go to the Wietzenb\"ock gauge. Applying Lorentz rotation (\ref{GrindEQ__36_}), one suppresses all the components of the inertial spin connection (\ref{DSspin}), then the proper tetrad acquires the form
 \begin{equation}
h^{a} {}_{\mu } =\left(
\begin{array}{cccc}
\left(1- {\textstyle \frac{1}{3}}\Lambda r^2 \right)^{1/2} & 0 & 0 & 0 \\
 0 & \frac{\sin \theta  \cos \varphi}{\left(1- {\textstyle \frac{1}{3}}\Lambda r^2 \right)^{1/2}} & r \cos \theta  \cos \varphi  & -r \sin \theta  \sin \varphi  \\
 0 & \frac{\sin \theta \sin \varphi}{\left(1- {\textstyle \frac{1}{3}}\Lambda r^2 \right)^{1/2}} & r \cos \theta \sin \varphi  & r \sin \theta \cos \varphi \\
 0 & \frac{\cos \theta}{\left(1- {\textstyle \frac{1}{3}}\Lambda r^2 \right)^{1/2}} & -r \sin \theta & 0 \\
\end{array}
\right)\,.
\m{protet_AdS}
\end{equation}
Because the components of the current $\scJ {}^\alf (\xi )$ are invariant under arbitrary local Lorentz rotations they are left the same for the proper tetrad (\ref{protet_AdS}) that is $\scJ {}^\alf (\xi )=0$.

\subsection{Discussion}
\m{discussion+}

Discussing the conserved current $\scJ {}^\alf (\xi )$ defined in (\ref{current_0}) for a freely falling observer, we turn to the both cases, the FLRW universe and the (A)dS space. We have found that in both the cases all its components are zero: $\scJ {}^\alf (\xi )=0$.  This means that such observers do not measure densities of energy and momentum, they are zero's in their own frame. It is quite natural because the observers really are frozen in the Hubble flow. Because our formalism is covariant the components $\scJ {}^\alf (\xi )$ for such freely falling observers in arbitrary coordinates continue to be vanished. These results show that our definitions of conserved currents and related superpotentials in TEGR are consistent and powerful.

It is quite constructive and instructive to compare our results with those in \cite{f(T)5} for the FLRW case. The authors under the requirement of a consistence of field equations in $f(T)$ theories are searching proper tetrads for solutions with more popular symmetries. Besides, they find appropriate inertial spin connections for diagonal tetrads. For the sake of definiteness we note that in \cite{f(T)5} the same metric element (\ref{metfrid}) and the same diagonal tetrad (\ref{tetradFLRW}) are under consideration.

In our study, the inertial spin connection (\ref{inertialSCFLRW}) and the proper tetrad (\ref{proptet_FLRW}) unite all the three possibilities of the curvature sign. In the case of spatially flat universe, $k=0$, we note that our inertial spin connection components (\ref{inertialSCFLRW}) related to the diagonal tetrad and our proper tetrad (\ref{proptet_FLRW}) are the same as in \cite{f(T)5}. For the case $k=+1$, they become complex ones, and there is no such a solution in \cite{f(T)5}. For the case $k=-1$,  (\ref{inertialSCFLRW}) and (\ref{proptet_FLRW}) are real and coincide exactly with the real solution for $k=-1$ in \cite{f(T)5}. Thus, we add \cite{f(T)5}.

On the other hand, the authors of \cite{f(T)5} present solutions that we cannot give. We derive here the components of the inertial spin connection related to the diagonal tetrad only. In the united form they are
\begin{equation}
\begin{array}{cccc}

\stackrel{\bullet}{A}{}^1{}_{22} =-\stackrel{\bullet}{A}{}^2{}_{12} =-\chi
~,~
\stackrel{\bullet}{A}{}^1{}_{33} =-\stackrel{\bullet}{A}{}^3{}_{13} =-\chi \sin\theta\,,
\\
\stackrel{\bullet}{A}{}^1{}_{23} =-\stackrel{\bullet}{A}{}^2{}_{13} =\sqrt{k} r \sin\theta~,~
\stackrel{\bullet}{A}{}^2{}_{31} =-\stackrel{\bullet}{A}{}^3{}_{21} ={\sqrt{k}}/{\chi}\,,
\\ ~,~
\stackrel{\bullet}{A}{}^1{}_{32} =-\stackrel{\bullet}{A}{}^3{}_{12} =-\sqrt{k} r ~,~
\stackrel{\bullet}{A}{}^2{}_{33} =-\stackrel{\bullet}{A}{}^3{}_{23} =-\cos\theta \,.
\m{SPIN}
\end{array}
\end{equation}
It turns out that calculation of the components of the current (\ref{current_0}) for the diagonal tetrad (\ref{tetradFLRW}), GR spin connection (\ref{friedlevispin}), proper vector (\ref{proper_FLRW}), but with inertial spin connection (\ref{SPIN}) instead of (\ref{inertialSCFLRW}), gives again  $\scJ {}^\alf (\xi )=0$.

Then we have 4 permissible variants of the inertial spin connection, (\ref{inertialSCFLRW}) and (\ref{SPIN}), uniting the cases $k=+1$ and $k=-1$, which give the acceptable result $\scJ {}^\alf (\xi )=0$. Among them 2 possibilities are complex ones. In \cite{f(T)5} the sense of complex spin connection (or complex proper tetrad) is not discussed. We try to do this now in the light of defining conserved quantities in TEGR. Already, we have remarked that an inertial spin connection plays an auxiliary role in the covariant TEGR. Indeed, it is absent in the field equations. The main requirement to the inertial spin connection is that a related curvature must be zero, it is the main requirement of the teleparallel approach. Thus, because the complex variants of (\ref{inertialSCFLRW}) and (\ref{SPIN}) have zero curvature they could be used for calculating conserved quantities in TEGR saving the real expressions for the diagonal tetrad. On the other hand, in any case one can choose the real variants from (\ref{inertialSCFLRW}) and (\ref{SPIN}) for the cases $k=-1$ and $k=+1$, respectively.

At last, let us note that the formalism developed in  \cite{f(T)5} does not permit to define a proper tetrad (and inertial spin connection related to diagonal tetrad) for the (A)dS space, whereas we define them here, they are (\ref{protet_AdS}) and (\ref{DSspin}).

\section{Concluding remarks}
\setcounter{equation}{0}
\m{Discussion}

From the start let us list main results presented in the present paper:

\noindent 1) In the framework of the covariant TEGR we have constructed conservation laws (\ref{DiffCL_1}) and (\ref{DiffCL_2}) with the conserved current (\ref{current_c_f}), including the gravitational energy-momen\-tum (\ref{E_M+}), and the superpotential (\ref{super}). All of these quantities are covariant with respect to coordinate transformations and are invariant with respect to local Lorentz rotations.

\noindent 2)  The local conserved quantities gives a possibility to present well defined conserved charges (\ref{ICQ_1}).

\noindent 3) In subsection \ref{Principles}, analyzing the structure of the conserved quantities we introduce the principle for determining inertial spin connection.

\noindent 4)  Discussing a philosophy of constructing conserving quantities in the covariant TEGR we compare it with constructing the covariantized both Einstein's pseudotensor and Freud's superpotential developed in \cite{KBL_1997}. We remark many analogous properties and some differences.

\noindent 5)  To show that our theoretical results are useful and powerful we provide some applications.
 First, we calculate mass of the Schwarzschild black hole, basing on observers at rest at spatial infinity, proper vectors of which coincide with the timelike Killing vector,  and obtain the acceptable result $M$. Second, we calculate densities of energy and momentum for freely falling observers in the FLRW universe and (A)dS space. We obtain zero quantities in all the cases that is quite acceptable because we consider observers at rest with respect to the Hubble flow.

The main mathematical tool of the present paper is the Noether theorem. Of course, already it has been used in teleparallel gravity. Thus, in \cite{[5],[6]}, using invariance with respect to action of specific groups, the Noether approach has been applied to find out analytical cosmological solutions in extended teleparallel gravities and to restrict potential variants of related Lagrangians. However, by this, there are no applications for constructing conserved quantities. In \cite{[4]}, it was used the diffeomorphism invariance of the Lagrangian, however, the final results have been presented in the simplest expressions, like (\ref{EM+}), with their problems. In \cite{Capozziello+2018}, the theories $f(R)$ and $f(T)$ are studied, but the Noether method uses invariance with respect to constant coordinate displacements only. Then, one obtains conserved pseudotensors only, not tensors.  Unlike the above, we give well structured expressions for conserved quantities with a clear and standard interpretation.

Applying the Noether theorem we have not excluded a
displacement vector $\xi^\alf$ from the consideration. Namely this allows us to construct local and
global conserved quantities trusting a covariance of both the kinds. Of course, there is no a contradiction between the new conservation law (\ref{DiffCL_2}) and the conservation law (\ref{EM+}) in the covariant TEGR \cite{Aldrovandi_Pereira_2013}, see Appendix A below.

 Besides of the aforementioned formal mathematical role of $\xi^\alf$, it has a quite principal role in constructing conserved quantities and their interpretation. For example, if a displacement vector is a timelike Killing vector at space infinity one can interpret a charge as a mass of a system, see section \ref{mass}; if $\xi^\alf$ is a proper vector of observer one can interpret components of the current as related densities, see section \ref{observer}; etc. This, classical approach differs fundamentally from many others used in works in teleparallel gravities, where authors identify an observer with a timelike tetrad vector, see, for example \cite{Obukhov+,[9],[10]} and references there in. Formally it looks permissible, however, there are problems in principles. Indeed, the observer is to be considered as an {\em external object} created for testing (observing) a physical or geometrical model. In contrast with this point of view, any tetrad vector components, being dynamical variables, are {\em internal objects} in TEGR, not an external structure.

The other very important question that must be discussed is the role of inertial spin connection, $\sA{}^i{}_{j\mu}$, in the covariant TEGR.  In spite of that the Moller Lagrangian (\ref{Moller}) and the TEGR Lagrangian (\ref{lag}) differs one from another, the field equations both in the Moller presentation (\ref{MollerEM}) and in the covariant TEGR presentation (\ref{EM}) are identical. Thus, the field equations do not contain $\sA{}^i{}_{j\mu}$ at all. Then, solutions to the equations (\ref{MollerEM}) and to (\ref{EM}) are the same. This means that on the level of equations there is no a necessity to define and introduce $\sA{}^i{}_{j\mu}$. The problems arise when one tries to define conserved quantities, and we resolve them here.

Last time a great attention arises to extended variants of teleparallel gravity. The most popular  is $f(T)$ theories, see \cite{f(T)5,Maluf1,Saridakis+,f(T)1,f(T)2,f(T)3,f(T)4,f(T)6} and references there in. Our principle of defining proper tetrad (the same, of defining $\sA{}^i{}_{j\mu}$ compatible to a given $\cA{}^i{}_{j\mu}$) is based on definition of conserved quantities in the covariant version of TEGR. It is impossible to introduce such a principle on the level of the field equations because their antisymmetric part is equal to zero identically. The other situation in $f(T)$ theories where the role of $\sA{}^i{}_{j\mu}$ becomes crucial one just in the system of equations. The inertial spin connection, $\sA{}^i{}_{j\mu}$, could be determined by its own field equations, which just coincide with the antisymmetric part of the field equations. It turns out that all modified teleparallel theories with second order field equations belong the same property, see \cite{f(T)7}.

The problem of determining both a proper tetrad and an inertial spin connection in a $f(T)$ gravity is too complicated to be solved in
general. In \cite{f(T)5,f(T)6}, it is suggested to study the problem on the basis of the symmetry considerations. The authors demonstrate the method in the case of axially and spherically symmetric spacetimes, homogeneous and isotropic spacetimes (FLRW universes) and maximally symmetric spacetimes ((anti-)de Sitter spaces).
Comparing our results with those in \cite{f(T)5}, we remark that, first, in the case of the Schwarzschild black hole they are identical, second, in the case of the FLRW universe they add one other, third, in the case of the (A)dS space we define a proper tetrad, whereas the authors \cite{f(T)5} do not.

In future, we plan to develop the presented here results to construct the Noether currents and superpotentials  1) both in $f(T)$ theories and other modifications of TEGR;  2) for perturbations in TEGR.

\appendix

\section{The TEGR conservation laws from metric GR }
\setcounter{equation}{0}
\m{Appendix}

Our study has been provided on the basis of the variational principle and with applying the Noether theorem. It is the standard and very economical way. It allows us to obtain a necessary result in very complicated cases and to give an appropriate interpretation. In this Appendix, we demonstrate that our main theoretical result, conservation laws (\ref{DiffCL_1}) and (\ref{DiffCL_2}) with the associated conserved quantities, can be derived {\em directly} starting from the usual form of the Einstein equations. As a result, first, we show that our way of varying the Lagrangian of covariant TEGR (\ref{lag}) is correct that  supports our position related to variation of divergence, see discussion in subsection \ref{FieldEquations}. Second, on the one hand,  we show that the inertial spin connection $\sA{}^i{}_{j\mu}$ is absent in the field equations in whole reflecting its auxiliary (external) character, on the other hand,  we show that $\sA{}^i{}_{j\mu}$ helps us to construct Lorentz covariant conserved quantities. Third, in independent way we demonstrate that conservation laws (\ref{DiffCL_1}) and (\ref{DiffCL_2}) are correct.

Let us derive the Einstein equations in the standard metric form:
\be
\cR_{\alf\beta}-\half g_{\alf\beta}\!\cR\, = \k \Theta_{\alf\beta}\,,
\m{Einstein}
\ee
where
\be
\cR_{\alf\beta}\, \equiv \di_\mu \cG{}^\mu{}_{\alf\beta} - \di_\alf \cG{}^\mu{}_{\mu\beta} + \cG{}^\mu{}_{\alf\beta}\cG{}^\nu{}_{\nu\mu} - \cG{}^\mu{}_{\nu\alf}\cG{}^\nu{}_{\mu\beta}; \qquad \cR \,\equiv g^{\alf\beta}\cR_{\alf\beta}
\m{Ricci_Scalar}
\ee
with $\cG{}^\mu{}_{\alf\beta}$ defined in (\ref{cG}). Contracting the equations (\ref{Einstein}) with $\sqrt{-g} h_a{}^\beta = hh_a{}^\beta$ and arising $\alf$ we rewrite them as
\be
h\l(\cR_a{}^{\alf}-\half h_a{}^\alf \cR\r) = \k h\Theta_a{}^{\alf}\,.
\m{Einstein_h}
\ee
It is evidently that the expressions in  (\ref{Einstein_h}) do not contain $\sA{}^i{}_{j\mu}$ at all, although (\ref{Einstein_h}) is Lorentz covariant in whole. Now, let us derive $\cG{}^\mu{}_{\alf\beta}$ in the new presentation:
\be
\cG{}^\mu{}_{\alf\beta} = \sG{}^\mu{}_{\alf\beta} - \sK{}^\mu{}_{\alf\beta}\,
\m{G_G_K}
\ee
that has been obtained from (\ref{contortion}) with making the use of (\ref{circ_A}) and (\ref{sG}).
In (\ref{G_G_K}) the left hand side is Lorentz invariant in whole although it does not contain $\sA{}^i{}_{j\mu}$ at all. Whereas the right hand side of (\ref{G_G_K}) is evidently Lorentz invariant in parts, $\sG{}^\mu{}_{\alf\beta}$ and $\sK{}^\mu{}_{\alf\beta}$, which contain $\sA{}^i{}_{j\mu}$.

Now let us substitute (\ref{G_G_K}) into (\ref{Einstein_h}) and provide identical transformations preserving the evident Lorentz covariance. After very prolonged but direct calculations one obtains
 \be
 - \kappa h \ssJ_a{}^\alf + \di_\rho \l(h \sS_a{}^{\alf\rho}\r) = \k h\Theta_a{}^\alf\,,
 \m{EM+L}
 \ee
where
\bea
\ssJ_a{}^\alf&\equiv & - \frac{h_a{}^\alf}{h}\sL + \frac{1}{\kappa} h_a{}^\pi\sS_c{}^{\alf\beta}\sT{}^c{}_{\pi\beta} +  \frac{1}{\kappa} \sA{}^c{}_{a\beta}\sS_c{}^{\alf\beta} \,,     \m{J+L}\\
 \sS_a{}^{\alf\rho} &\equiv & \l(\half  {\sT}_\beta{}^{\alf\rho} -  {\sT}{}^{[\alf\rho]}{}_\beta - 2{\sT}{}^{\pi[\alf}{}_\pi \delta^{\rho]}_\beta  \r)h_a{}^\beta\,. \m{S+L}
 \eea
 We note that to achieve (\ref{EM+L}) we have used the definitions: (\ref{sG}),  (\ref{contor}), (\ref{tor}) and (\ref{lag}), and the identity $\sR{}^\mu{}_{\nu\alf\beta} \equiv 0$. Comparing (\ref{EM+L}) with (\ref{EM}), (\ref{J+L}) with (\ref{J}) and (\ref{S+L}) with (\ref{S}) we conclude that the equations  (\ref{EM+L}) are exactly the equations (\ref{EM+}). By the above we have shown that 1) variation of the TEGR Lagrnagian (\ref{lag}) is correct; 2) equations (\ref{EM})) are, indeed, equivalent to the Einstein equations (\ref{Einstein}); 3) the inertial spin connection $\sA{}^i{}_{j\mu}$ is absent in the field equations (\ref{EM}) in whole demonstrating the auxiliary character of $\sA{}^i{}_{j\mu}$.

 Now, we will show that (\ref{DiffCL_1}) and (\ref{DiffCL_2}) can be derived from (\ref{EM+L}) (this means from (\ref{Einstein})) by identical transformations only. Let us rewrite (\ref{EM+L}) in the form:
 \be
 \kappa h \ssJ_a{}^\alf + \k h\Theta_a{}^\alf = \di_\rho \l(h \sS_a{}^{\alf\rho}\r)  \,.
 \m{EM+L+}
 \ee
Here, substituting (\ref{J+L}) with taking into account (\ref{tor}) one obtains
 \be
 h_a{}^\pi\l[ h\sS_c{}^{\alf\rho} \l( \cN_\pi h^c{}_{\rho} - \cN_\rho h^c{}_{\pi} + \sA{}^c{}_{b\pi}h^b{}_\rho\r) - \kappa\sL \delta^\alf_\pi + \kappa h \Theta_\pi{}^\alf\r] = \di_\rho \l( h \sS_a{}^{\alf\rho}\r)   \,.
 \m{comparison_1}
 \ee
 Contracting it with $h^a{}_{\sig}\xi^\sig$, replacing  the term $h\sS_a{}^{\alf\rho} \xi^\sig \cN_\rho h^a{}_{\sig}$ to the right  hand side and adding $h \sS_a{}^{\alf\rho}  h^a{}_{\sig}\cN_\rho\xi^\sig$ to both sides we obtain finally
 \bea
 &&\l[ h\sS_c{}^{\alf\rho} \l( \cN_\sig h^c{}_{\rho} + \sA{}^c{}_{b\sigma}h^b{}_\rho\r) - \kappa\sL \delta^\alf_\sig + \kappa h \Theta_\sig{}^\alf\r]\xi^\sigma + h \sS_a{}^{\alf\rho}  h^a{}_{\sig}\cN_\rho\xi^\sig \m{comparison}\\ && = \di_\rho \l( h \sS_a{}^{\alf\rho}\r)  h^a{}_{\sig}\xi^\sig +   h\sS_a{}^{\alf\rho} \xi^\sig \cN_\rho h^a{}_{\sig} + h \sS_a{}^{\alf\rho}  h^a{}_{\sig}\cN_\rho \xi^\sig= \cN_\rho \l( h \sS_a{}^{\alf\rho}  h^a{}_{\sig}\xi^\sig \r)\,.
 \nonumber
 \eea
One easily recognizes that (\ref{comparison}) is the conservation law (\ref{DiffCL_2}). One can obtain (\ref{DiffCL_1}) after differentiating (\ref{comparison}) as well.

 By deriving (\ref{comparison}) directly from (\ref{Einstein}) we show 1) that there is no a contradiction between the new conservation law (\ref{DiffCL_2}) and the conservation law (\ref{EM+}) in the framework of the covariant TEGR; 2) that with using $\sA{}^i{}_{j\mu}$ we have a possibility to construct Lorentz covariant conserved quantities, which allow us to construct well defined charges (\ref{ICQ_1}). At last, 3) we present the independent way in constructing conservation laws (\ref{DiffCL_1}) and (\ref{DiffCL_2}).

\bigskip

\noindent {\bf Acknowledgments.} The authors acknowledge the support from the Program of development of M.V. Lomonosov Moscow State University
(Leading Scientific School 'Physics of stars, relativistic objects and galaxies'). A. T. is supported
by the Russian Government Program of Competitive Growth of Kazan Federal University. Authors are grateful to Martin Kr{\v s}{\v s}{\'a}k,  Laur J\"arv
and Manuel Hohmann
for discussions.

\ed